\documentclass[a4paper,fleqn,usenatbib]{mnras}

\usepackage{mathptmx}
\usepackage{txfonts}
\hypersetup{draft}

\usepackage[T1]{fontenc}
\usepackage{ae,aecompl}

\usepackage{graphicx}   
\usepackage{amssymb}    

\title[Molecular abundances in 67P]{Evolution of CO$_2$, CH$_4$, and OCS abundances relative to H$_2$O in the coma of comet 67P around perihelion from Rosetta/VIRTIS-H observations}

\author[D. Bockel\'ee-Morvan et al.]{Dominique Bockel\'ee-Morvan,$^{1}$\thanks{E-mail: dominique.bockelee@obspm.fr}
J. Crovisier,$^{1}$
S. Erard,$^{1}$
F. Capaccioni,$^{2}$
C. Leyrat,$^{1}$
\newauthor
G. Filacchione,$^{2}$
P. Drossart,$^{1}$
T. Encrenaz, $^{1}$
N. Biver,$^{1}$
M.-C. de Sanctis,$^{2}$
\newauthor
B. Schmitt,$^{3}$
E. K\"{u}hrt,$^{4}$
M.-T. Capria,$^{2}$
M. Combes,$^{1}$
M. Combi,$^{5}$
N. Fougere,$^{5}$
\newauthor
G. Arnold,$^{4}$
U. Fink,$^{6}$ 
W. Ip,$^{7}$ 
A. Migliorini,$^{2}$
G. Piccioni,$^{2}$
G. Tozzi$^{8}$
\\
$^{1}$ LESIA, Observatoire de Paris, PSL Research University, CNRS, Sorbonne Universit\'es, UPMC Univ. Paris 06, \\
Univ. Paris-Diderot, Sorbonne Paris Cit\'e, 5 place Jules Janssen, 92195 Meudon, France\\
$^{2}$ INAF-IAPS, Istituto di Astrofisica e Planetologia Spaziali, via del fosso del Cavaliere, 100, 00133, Rome, Italy \\
$^{3}$ Universit\'e Grenoble Alpes, CNRS, Institut de Plan\'etologie et d'Astrophysique de Grenoble, Grenoble, France  \\
$^{4}$ Institute for Planetary Research, Deutsches Zentrum f\"{u}r Luft- und Raumfahrt (DLR), Berlin, Germany \\
$^{5}$ Department of Climate and Space Sciences and Engineering, University of Michigan, Ann Arbor, Michigan, 48109, USA \\
$^{6}$ Lunar Planetary Laboratory, University of Arizona, Tucson, USA \\
$^{7}$ National Central University, Taipei, Taiwan \\
$^{8}$ INAF, Osservatorio Astrofisico di Arcetri,  Largo E. Fermi 5, 50125 Firenze, Italy
}

\date{Accepted XXX. Received YYY; in original form ZZZ}

\pubyear{2016}

\begin{document}
\label{firstpage}
\pagerange{\pageref{firstpage}--\pageref{lastpage}}
\maketitle

\begin{abstract}
Infrared observations of the coma of 67P/Churyumov-Gerasimenko
were carried out from July to September 2015, i.e., around
perihelion (13 August 2015), with the high-resolution channel of
the VIRTIS instrument onboard Rosetta. We present the analysis
of fluorescence emission lines of H$_2$O, CO$_2$, $^{13}$CO$_2$,
OCS, and CH$_4$ detected in limb sounding with the field of view
at 2.7--5 km from the comet centre. Measurements are sampling
outgassing from the illuminated southern hemisphere, as revealed
by H$_2$O and CO$_2$ raster maps, which show anisotropic
distributions, aligned along the projected rotation axis. An abrupt increase
of water production is observed six days after perihelion. In the mean time,
CO$_2$, CH$_4$, and OCS abundances relative to water increased by
a factor of 2 to reach mean values of 32 \%, 0.47 \%, and 0.18\%,
respectively, averaging post-perihelion data. We interpret these
changes as resulting from the erosion of volatile-poor surface
layers. Sustained dust ablation due to the
sublimation of water ice maintained volatile-rich layers near the
surface until at least the end of the considered period, as
expected for low thermal inertia surface layers. The large
abundance measured for CO$_2$ should be representative of the 67P
nucleus original composition, and indicates that 67P is a
CO$_2$-rich comet. Comparison with abundance ratios measured in
the northern hemisphere shows that seasons play an important role
in comet outgassing. The low CO$_2$/H$_2$O values measured above
the illuminated northern hemisphere are not original, but the
result of the devolatilization of the uppermost layers.


\end{abstract}




\section{Introduction}

Comets are among the most pristine objects of the Solar System. The chemical composition of nucleus ices should provide insights for the conditions of formation and evolution of the early Solar System. A large numbers of molecules have now been identified in cometary atmospheres, both from ground-based observations and space, including in situ investigations of cometary atmospheres \citep{dbm2004,Cochran2015,Leroy2015,Biver2015a,Biver2016}. Molecular abundances relative to water present strong variations from comet to comet, and also vary along comet orbit \citep[e.g., ][]{Biver2016,Ootsubo2012,McKay2015,McKay2016}. This chemical diversity is often interpreted as reflecting different formation conditions in the primitive solar nebula \citep[e.g., ][]{Ahearn2012}. However, questions arise concerning the extent to which abundances measured in cometary atmospheres are representative of the pristine composition of nucleus ices. Models investigating the thermal evolution and outgassing of cometary nuclei show that the outgassing profiles of cometary molecules depend on numerous factors such as molecule volatility, thermal inertia of the nucleus material, nature of water ice structure, and dust mantling \citep{deSanctis2005,deSanctis2010a,deSanctis2010b,Marboeuf2014}.

The study of the development of cometary activity, with the goal
to relate coma and nucleus chemical properties, is one of the main
objectives of the Rosetta mission  of the European Space Agency
\citep{Schulz2012}. Rosetta reached comet
67P/Churyumov-Gerasimenko (hereafter referred to as 67P) in August
2014 at $r_h$ = 3.5 AU from the Sun, and accompanied it in its
journey towards perihelion (13 August 2015, $r_h$ = 1.24 AU), with
the end of the mission in September 2016. The scientific
instruments on the orbiter and Philae lander provided
complementary information on the physical and chemical properties
of the nucleus surface and subsurface, and of the inner coma. They
revealed a dark, organic-rich and low-density bi-lobed nucleus
with a morphologically complex surface showing different
geological terrains, some of them with smooth dust-covered areas
\citep{Sierks2015,Capaccioni2015,Patzold2016}. Dust jets
originating from active cliffs, fractures and pits were observed
\citep{Vincent2016}. Water vapour was first detected in June 2014,
at 3.93 AU from the Sun \citep{Gulkis2014,Gulkis2015}. In data
obtained pre-equinox (May 2015), the water outgassing was found to
correlate well with solar illumination, though with a large excess
from the brighter and bluer Hapi region situated in the northern
hemisphere
\citep{Biver2015b,dbm2015,Feldman2015,Migliorini2016,Fink2016,Fougere2016a}.
In the Hapi active region, VIRTIS has observed that surficial
water ice has diurnal variability showing sublimation and
condensation cycle occurring with the change of illumination
conditions \citep{deSanctis2015}. In contrast, CO$_2$ 
originated mainly from the poorly illuminated southern hemisphere
\citep{Haessig2015,Migliorini2016,Fink2016,Fougere2016a}, where
VIRTIS identified a CO$_2$ ice-rich area \citep[specifically in
the Anhur region][]{Filacchione2016a}.

The large obliquity (52$^{\circ}$) of the 67P rotation axis \citep{Sierks2015} leads
to strong seasonal effects on its nucleus, with the northern
regions experiencing a long summer at large distances from the Sun
whereas the southern polar regions are subject to a short-lived,
but extremely intense summer season around perihelion. In this
paper, we present observations of H$_2$O, CO$_2$, $^{13}$CO$_2$,
OCS, and CH$_4$ in the vapour phase undertaken with the high spectral
resolution channel of the Visible InfraRed Thermal Imaging
Spectrometer (VIRTIS) \citep{Coradini2007} near perihelion (early
July to end of September 2015). The paper is organized as follow:
Section~\ref{sec:obs} presents the VIRTIS-H instrument, the data
set and raster maps of H$_2$O and CO$_2$ which provide the context
for the off-limb observations studied in the paper; in
Sect.~\ref{sec:model}, we present the detected molecular emission
lines and the fluorescence models used for their analysis;
section~\ref{sec:results} presents the data analysis and the
column densities measured for H$_2$O, CO$_2$, CH$_4$, and OCS, and
their relative abundances; an interpretation of the results and a
discussion follows in Sect.~\ref{sec:discussion}.

\section{Observations}
\label{sec:obs}

\subsection{VIRTIS-H instrument}
\label{sec:VH}

\begin{figure*}
    \includegraphics[width=15cm]{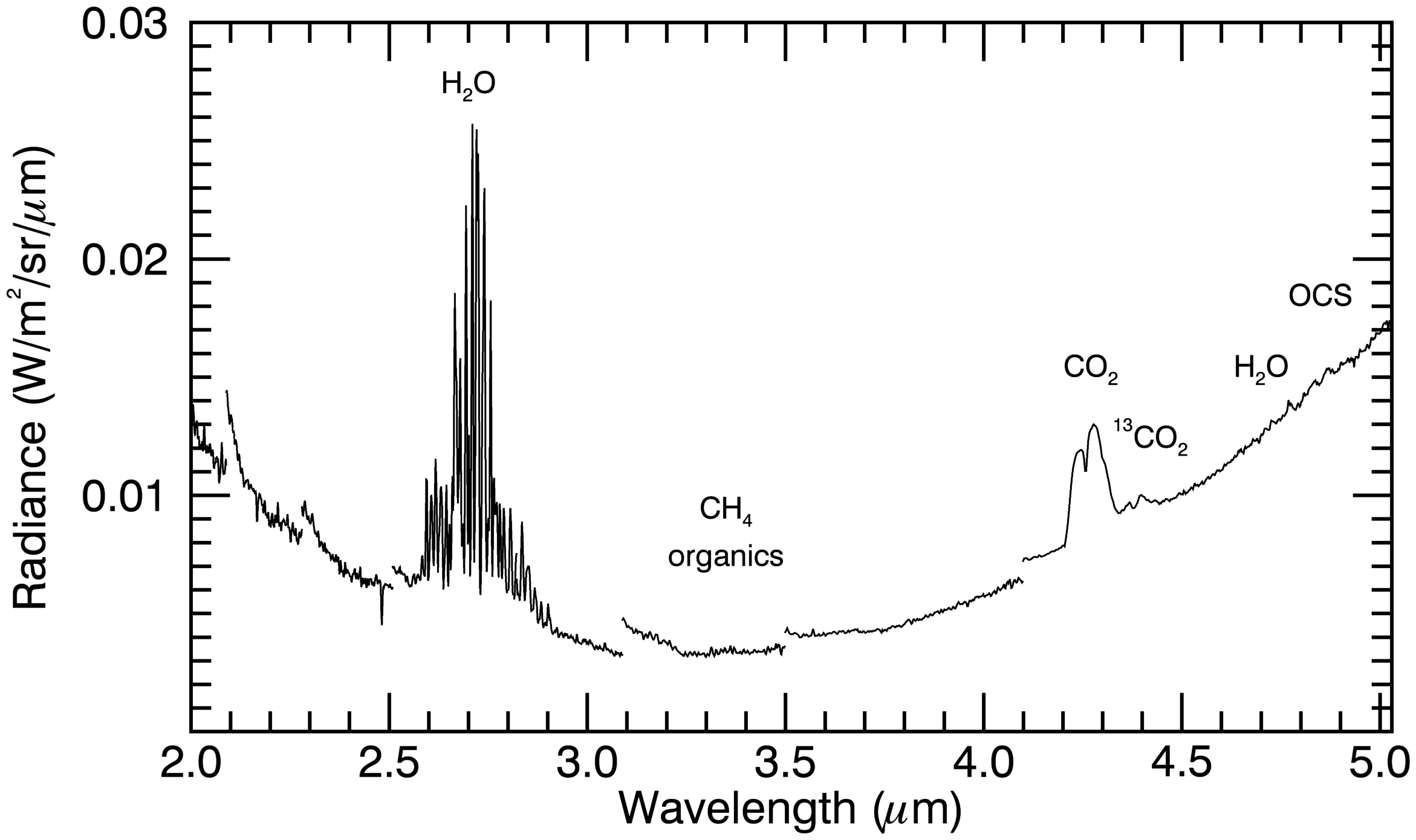}
   \caption{Full VIRTIS-H spectrum obtained by averaging the 3 data cubes
   obtained on 19 Aug. 2015. At this scale, the faint emissions from CH$_4$, OCS and H$_2$O 4.5--5 $\mu$m bands
   are not easily distinguishable. The short wavelength range of the spectral orders is contaminated by instrumental straylight (e.g., at 2.1 and 2.28 $\mu$m), resulting in a higher radiance.  }
    \label{fig:fullspectrum}
\end{figure*}

The VIRTIS instrument is composed of two
channels: VIRTIS-M, a spectro-imager operating both in the visible
(0.25--1 $\mu$m) and infrared  (1--5 $\mu$m) ranges at moderate
spectral resolution ($\lambda$/$\Delta \lambda$ = 70-380), and
VIRTIS-H, a point spectrometer in the infrared (1.9--5.0
$\mu$m) with higher spectral resolution capabilities \citep{Coradini2007}. The infrared channel of VIRTIS-M
stopped providing cometary data in May 2015, due to a cryocooler failure. Only VIRTIS-H data are used in this paper.

VIRTIS-H is a cross-dispersing spectrometer consisting of a telescope, an entrance
slit, followed by a collimator, and a prism separating eight
orders of a grating \citep{Drossart2000,Coradini2007}. The spectra investigated in this paper
are in orders 0 (4.049--5.032 $\mu$m) and 2 (3.044--3.774 $\mu$m).
Each order covers 432 $\times$ 5 pixels on the detector. A "pixel map" locates the illuminated pixels.
The field of view corresponds to a 5-pixel wide area on the focal plane array;
in nominal mode (most data), the summation of the signals received on these
pixels (actually only the 3 more intense) is performed on-board. The
spectra are therefore sampled by 3456 spectral elements, 432 in each
grating order. As for most Rosetta instruments, the line of sight of VIRTIS is along
the Z-axis of the spacecraft (S/C). The instantaneous field of view (FOV) of the
VIRTIS-H instrument is 0.58 $\times$ 1.74 mrad$^2$ (the larger
dimension being along the Y axis).

The VIRTIS-H data are processed using the so-called CALIBROS
pipeline. This pipeline subtracts dark current, corrects for
the instrumental function (odd/even pixels effects,
de-spiking), and calibrates the absolute flux of the spectra in
radiance (W/m$^2$/sr/$\mu$m). Stray light is present in some spectra on
the short-wavelength side of each order, depending on the location of the
nucleus with respect to the slit (Fig~\ref{fig:fullspectrum}).

The nominal numerical spectral resolving power $\lambda$/$\Delta\lambda$ varies between
1300 and 3500 within each order, where $\Delta\lambda$ is the grating resolution. The spectral sampling interval is 1.2--1.3 $\Delta\lambda$, and thus above Nyquist sampling. In the current calibration, odd/even effects were overcome by performing spectral smoothing with a boxcar average over 3 spectral channels, thereby  reducing the
effective spectral resolution to a value $\sim$ 2.7 $\times$ $\Delta\lambda$, which was experimentally verified by fitting spectra of the H$_2$O and CO$_2$ bands in order 4 and 0, respectively \citep{Debout2015}. The effective spectral resolution with the current calibration is $\sim$ 800 at 3.3 and 4.7 $\mu$m.

The geometric parameters of the observations have been computed
using nucleus shape model SHAP5 v1.1, derived from OSIRIS images \citep{Jorda2016}.

\subsection{Data set}
\label{sec:dataset}
\begin{table*}
    \caption{Log of VIRTIS-H observations.}
    \label{tab:log}
    \begin{tabular}{lccccccclcc} 
        \hline
Obs Id & Start time & Duration & Date wrt& $r_h$ & $\Delta$(S/C)   & phase & \multicolumn{2}{c}{Pointing$^a$} & \multicolumn{2}{c}{South pole$^b$}\\
&&& perihelion &&&& $\rho$& $PA$ & $PA$ & Aspect\\
& (UT) & (hr) & (d) & (AU) & (km)  & ($^{\circ}$) & (km) &  ($^{\circ}$)& ($^{\circ}$) & ($^{\circ}$) \\
        \hline
   00395011055 & 2015-07-08T21:11:00.3 &  3.51 &  -35.2 & 1.316 &   150 &    90 &  2.82 &  259 &   328 &   108 \\
   00395322550 & 2015-07-12T11:46:52.6 &  3.66 &  -31.6 & 1.302 &   153 &    89 &  2.70 &  274 &   327 &   101 \\
   00395742154 & 2015-07-17T08:15:55.9 &  2.56 &  -26.7 & 1.286 &   177 &    90 &  3.10 &  272 &   302 &   137 \\
   00396199623 & 2015-07-22T15:23:57.9 &  4.27 &  -21.4 & 1.271 &   169 &    89 &  2.69 &  266 &   319 &   110 \\
   00396220410 & 2015-07-22T21:10:24.7 &  3.60 &  -21.2 & 1.270 &   170 &    89 &  2.98 &  266 &   319 &   109 \\
   00396659230 & 2015-07-27T23:06:40.5 &  3.54 &  -16.1 & 1.259 &   177 &    90 &  4.01 &  247 &   311 &    60 \\
   00396826054 & 2015-07-29T21:20:59.8 &  3.24 &  -14.2 & 1.255 &   179 &    90 &  3.56 &  265 &   286 &    44 \\
   00396842044 & 2015-07-30T01:47:25.8 &  3.91 &  -14.0 & 1.255 &   177 &    90 &  3.59 &  265 &   281 &    43 \\
   00397038715 & 2015-08-01T08:25:16.8 &  0.67 &  -11.7 & 1.252 &   214 &    90 &  3.08 &  270 &   238 &    52 \\
   00397871165 & 2015-08-10T23:32:29.0 &  2.80 &   -2.1 & 1.244 &   323 &    89 &  6.23 &  286 &   242 &   123 \\
   00398347798 & 2015-08-16T11:56:25.1 &  3.31 &    3.4 & 1.244 &   327 &    89 &  3.58 &  267 &   288 &   127 \\
   00398603680 & 2015-08-19T11:08:06.5 &  3.24 &    6.4 & 1.246 &   329 &    90 &  2.96 &  263 &   302 &   115 \\
   00398619671 & 2015-08-19T15:34:33.4 &  3.91 &    6.6 & 1.246 &   327 &    90 &  2.97 &  262 &   303 &   114 \\
   00398640452 & 2015-08-19T21:20:58.5 &  3.24 &    6.8 & 1.246 &   326 &    90 &  3.47 &  262 &   304 &   112 \\
   00398729980 & 2015-08-20T22:13:06.6 &  3.91 &    7.8 & 1.247 &   325 &    90 &  4.58 &  258 &   307 &   105 \\
   00398970868 & 2015-08-23T17:07:54.5 &  3.10 &   10.6 & 1.250 &   340 &    87 &  3.78 &  256 &   309 &    80 \\
   00398986563 & 2015-08-23T21:29:29.5 &  3.78 &   10.8 & 1.251 &   343 &    87 &  3.82 &  255 &   308 &    78 \\
   00399198695 & 2015-08-26T08:17:52.6 &  1.02 &   13.3 & 1.254 &   414 &    83 &  4.25 &  264 &   302 &    59 \\
   00399208316 & 2015-08-26T10:58:13.4 &  3.56 &   13.4 & 1.254 &   410 &    83 &  3.95 &  263 &   301 &    57 \\
   00400195914 & 2015-09-06T21:25:16.9 &  3.91 &   24.8 & 1.280 &   339 &   108 &  4.92 &  266$^c$ &   231 &   124 \\
   00400216374 & 2015-09-07T03:06:16.9 &  3.24 &   25.0 & 1.281 &   332 &   110 &  4.87 &  268$^c$ &   231 &   127 \\
   00400232094 & 2015-09-07T07:28:16.8 &  3.91 &   25.2 & 1.282 &   327 &   111 &  4.83 &  268$^c$ &   232 &   131 \\
   00400252549 & 2015-09-07T13:09:16.0 &  3.24 &   25.5 & 1.282 &   322 &   112 &  4.83 &  270$^c$ &   232 &   134 \\
   00400407713 & 2015-09-09T08:15:16.0 &  1.08 &   27.3 & 1.288 &   331 &   119 &  3.14 &  263 &   255 &   156 \\
   00400433767 & 2015-09-09T15:29:34.0 &  4.05 &   27.6 & 1.289 &   324 &   120 &  3.72 &  262 &   263 &   157 \\
   00400470243 & 2015-09-10T01:37:26.0 &  4.18 &   28.0 & 1.290 &   318 &   120 &  5.78 &  263$^d$ &   274 &   158 \\
   00400748481 & 2015-09-13T06:54:44.2 &  3.51 &   31.2 & 1.301 &   313 &   110 &  3.58 &  265 &   307 &   132 \\
   00400765976 & 2015-09-13T11:46:19.3 &  3.64 &   31.4 & 1.302 &   312 &   108 &  3.61 &  265 &   307 &   129 \\
   00401717336 & 2015-09-24T11:55:21.0 &  4.07 &   42.4 & 1.347 &   488 &    63 &  5.02 &  275 &   317 &    60 \\
   00401994267 & 2015-09-27T16:57:54.5 &  3.51 &   45.6 & 1.362 &  1057 &    51 &  6.88 &  282 &   329 &    35 \\
        \hline
    \end{tabular}

    {\raggedright
    $^a$ Except when indicated, the FOV is staring at a fixed distance $\rho$ from the comet centre at the given position angle $PA$ which is close to the Sun direction ($PA$(Sun) = 270$^{\circ}$).

 $^b$ Orientation of South Pole direction defined by its position angle in the plane of the sky and its aspect angle with respect to the line of sight.

 $^c$ The S/C was continuously slewing along the $Y$ axis, i.e.,  perpendicularly to the comet-Sun line providing data at $y$=--4 to + 4 km with $x$ fixed; $PA$ was ranging from 220--310$^{\circ}$, only the mean $PA$ is indicated.

 $^d$ Two stared positions were commanded, whereas we used the closest to comet center.

 }

\end{table*}

The data set was selected with the objective of studying spectral
signatures of a number of species and derive their relative
abundances. As presented in the next sections (see
Fig.~\ref{fig:fullspectrum}), five molecules (H$_2$O, CO$_2$,
$^{13}$CO$_2$, OCS, and CH$_4$) are detected unambiguously around
perihelion (on 13.0908 August UT), not considering organic
molecules contributing to the 3.4 $\mu$m band. Whereas H$_2$O and
CO$_2$ present strong signatures at 2.7 and 4.3 $\mu$m,
respectively, which were detected by VIRTIS-H starting October 2014
\citep{dbm2015}, these bands became optically thick around
perihelion time. Hence, in this paper we consider the July to September 2015
period during which optically thin H$_2$O and CO$_2$ hot-bands,
and $^{13}$CO$_2$ are detected. An appropriate radiative transfer
analysis \citep[following, e.g.,][]{Debout2016} of the optically
thick H$_2$O and CO$_2$ bands will be presented in forthcoming
papers. Specifically, we analysed the $\nu_3$-$\nu_2$ and
$\nu_1$--$\nu_2$ H$_2$O hot bands falling in the 4.4--4.5 $\mu$m
range, and the CO$_2$  $\nu_1$+$\nu_3$--$\nu_1$ bands at 4.3
$\mu$m.

Coma observations with VIRTIS-H were performed in various pointing
modes, e.g., stared pointing at a fixed limb distance in inertial
frame or raster maps. In this paper, we consider mostly those with
the FOV staring at a distance of less than 5 km from the comet
surface (distance from comet center less than 7 km) along the
projected comet-Sun line. These observations are those providing
the largest signal-to-noise ratios in the spectra. In addition, we
considered four data cubes with acquisitions scanning a line of
8-km long perpendicular to the Sun direction and above the
illuminated nucleus (data of 6--7 September 2015,
Table~\ref{tab:log}). The mean distances to the comet center ($\rho$) for the
considered data cubes are between 2.7 and 6.9 km
(Table~\ref{tab:log}).

Altogether thirty data cubes, lasting typically 3--4 h each, were
selected. They are listed in Table~\ref{tab:log}, together with
geometric parameters (phase angle, heliocentric distance $r_h$,
and distance of the S/C to the comet, mean distance of the FOV with respect to
the comet center and position angle). For these observations, the acquisition time is 3 s, except for the July 12th and 27th observations for which it is 1 s, and the dark rate is 4 (one dark image every 4 acquisitions). Frame summing has been performed on-board for data volume considerations, e.g., by summing pair of acquisitions. The total number of acquisitions in the data files is between 256 (case of first Aug. 26th observation) and 7424 (observation of July 27th).  

For most of the selected observations the
S/C was in the terminator plane (phase angle $\sim$ 90$^{\circ}$).
Therefore these observations were favourable for sampling
molecules outgassed from the illuminated regions of the nucleus.
However, the two last data cubes were acquired with a smaller
phase angle $\sim$ 50--60$^{\circ}$, with the closest points to
the line of sight (LOS)  situated from less illuminated regions,
so that less intense signals are expected (and indeed observed).
During the period, the Rosetta spacecraft was at distances from
the comet from 150 to 480 km (with the exception of the 27
September observation with the S/C at 1057 km,
Table~\ref{tab:log}). The VIRTIS-H FOV then ranged from 90
$\times$ 260 m to 280 $\times$ 850 m in the orthogonal plane containing the comet, which can be compared to the
typical size of 67P nucleus of 2 km equivalent radius. This means
that molecules detected in the field of view are potentially
originating from a broad range of longitudes and latitudes.


\subsection{Context VIRTIS-H raster maps}
\label{sec:context}
\begin{figure}
\hspace{0.15cm}\includegraphics[width=7.9cm]{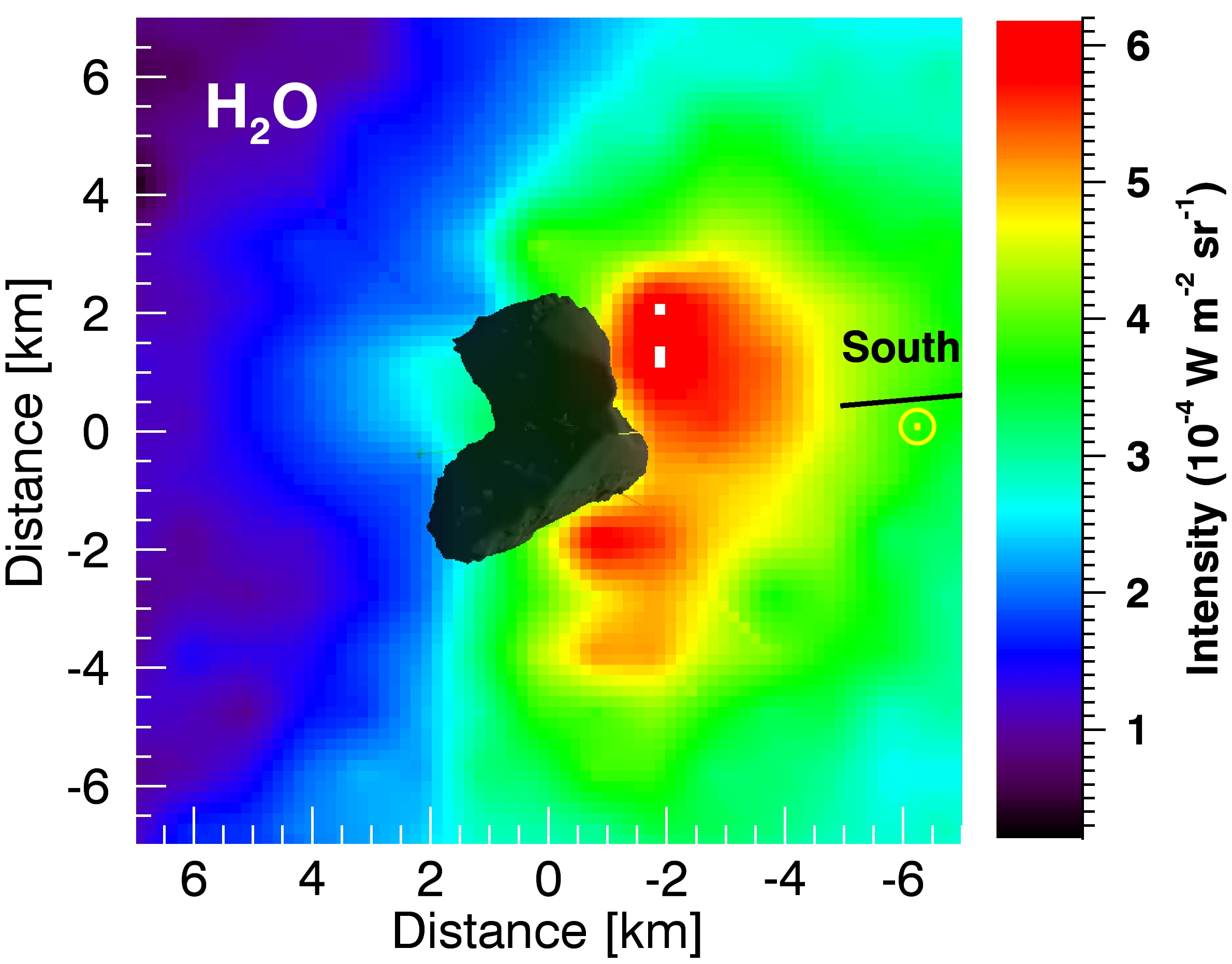}
    \includegraphics[width=8.0cm]{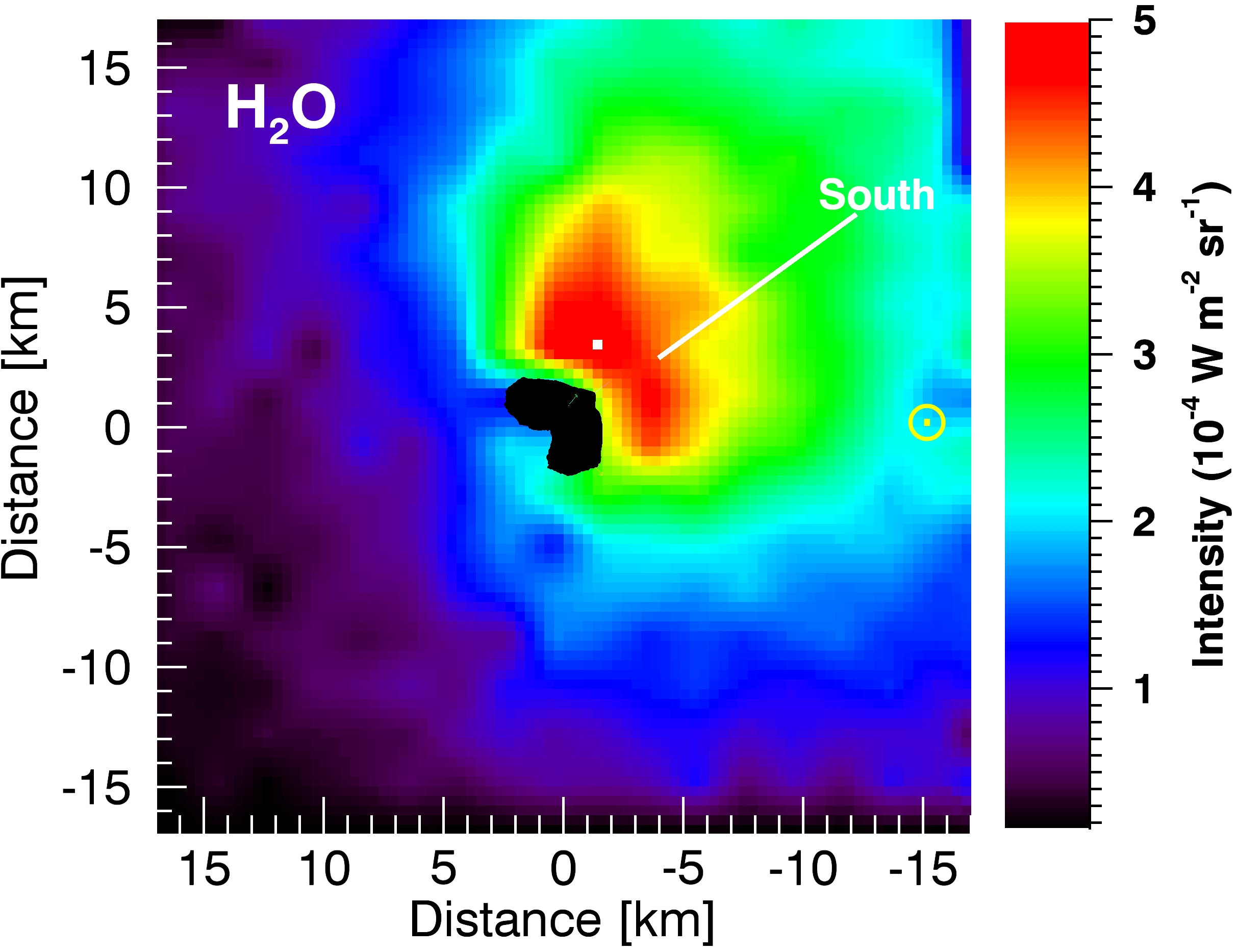}
    \includegraphics[width=8.20cm]{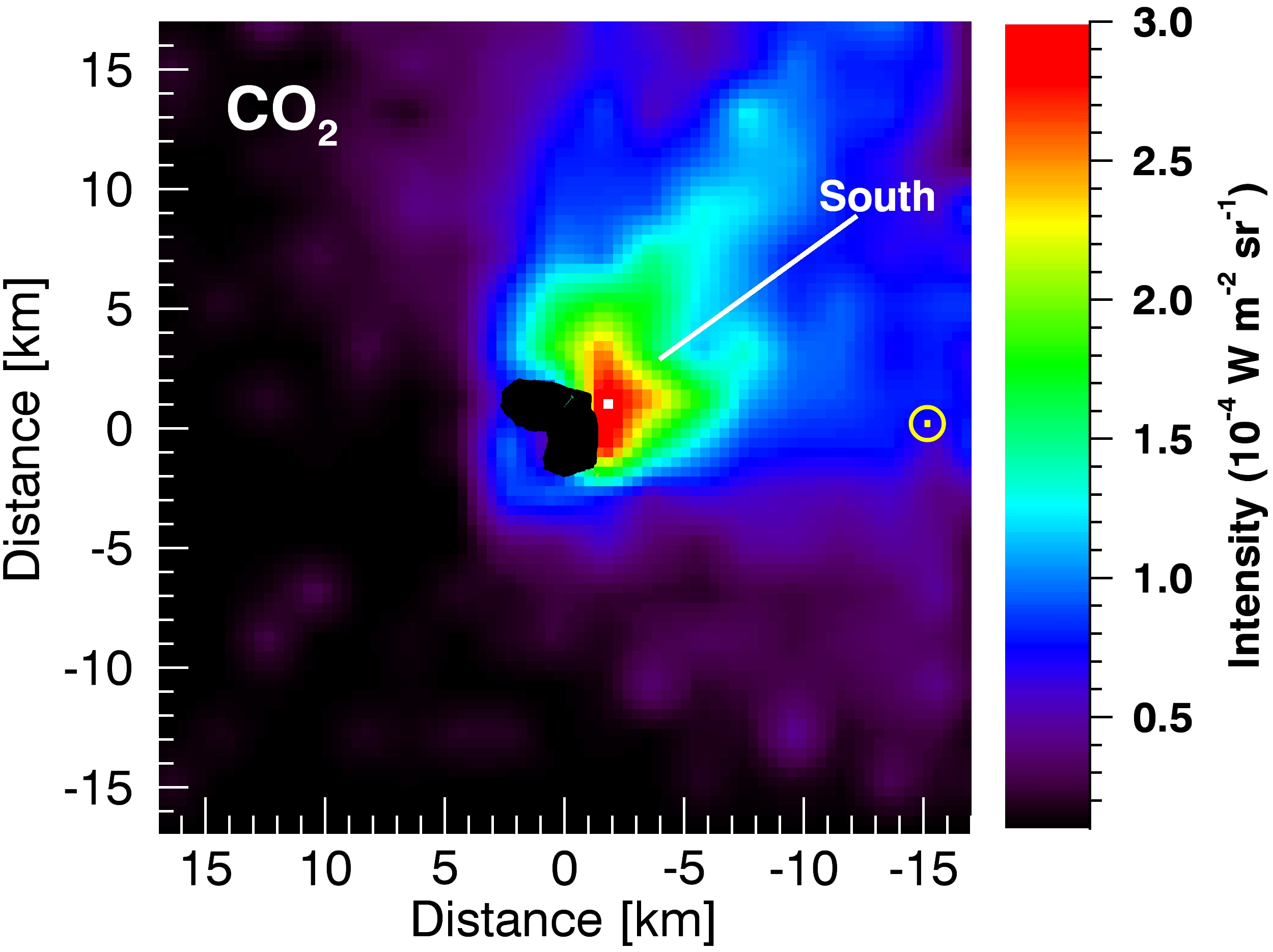}
   \caption{VIRTIS-H raster maps of the H$_2$O 2.7 and CO$_2$ 4.3 $\mu$m bands. Top: H$_2$O on 30 July 2015, 19h:25 UT (ObsId: 00396898661). Middle and bottom: H$_2$O and CO$_2$, respectively, on 9 August 2015, 3h:10 UT (ObsId: 00397703949). The projection of the rotation axis  (South direction) and Sun direction are indicated. The observational parameters are: line spacing of 0.6 km and 2 km, slew rate of 1.0 and 1.75 deg/min, for 30 July and 9 August, respectively. The maps were obtained after averaging spectra within cells of sizes of 1 km (30 July) and 2 km (9 August). }
    \label{fig:map}
\end{figure}

As a result of the ~52$^{\circ}$ obliquity and the orientation of the 67P rotation axis
\citep{Sierks2015}, the perihelion period corresponds to the time
where only the southern regions are illuminated. During the
July-September period, the sub-solar latitude varied between --26
$^{\circ}$ to --52$^{\circ}$, with the lowest value reached
on 4 September 2015.

In order to put the limb observations analysed in
this paper  into context, we show in Fig.~\ref{fig:map} VIRTIS-H raster maps of
the H$_2$O 2.7 $\mu$m and CO$_2$ 4.3 $\mu$m bands obtained on 30
July and 9 August 2015,  when Rosetta was at 181 km and 303 km from the comet, respectively. These maps (not listed in Table~\ref{tab:log}) were obtained by slewing the
spacecraft along the $X$ S/C axis (with the Sun being in the -$X$
direction), setting a line spacing in the $Y$ direction. The
observation duration is 3--4 h, i.e., 1/4 to 1/3 of the rotation
period of $\sim$ 12 h \citep{Sierks2015}, so there is some
smearing related to nucleus rotation. Spatial binning was applied,
by averaging spectra into cells of size of 1 and 2 km. After
baseline removal, the CO$_2$ band area was determined by gaussian
fitting, whereas that of the H$_2$O band by summing the radiances
measured in the wavelength interval 2.6--2.73 $\mu$m.

The distributions of the H$_2$O and
CO$_2$ band intensities display a fan-shaped morphology aligned along the South direction of the rotation axis (Fig.~\ref{fig:map}), indicating that these molecules were mainly released from the southern regions. Because the Rosetta S/C was moving around the comet, the position angle $PA$ of the rotation axis in the ($X$, $Y$) S/C frame changed with time, spanning value between 240 and 330$^{\circ}$ in the July-November timeframe (Table~\ref{tab:log}), with the Sun at +270$^{\circ}$  in the adopted $PA$ definition ($PA$ = 0 for the +$Y$ axis, $PA$ = 90$^{\circ}$ for +$X$). VIRTIS-H maps acquired from August to November 2015 in different $PA$ configurations all show a tight correlation between the orientation of the H$_2$O and CO$_2$ fans and the North-South line \citep{dbm2016ESLAB}. This is consistent with in situ data obtained by the Rosetta Orbiter
Spectrometer for Ion and Neutral Analysis (ROSINA) \citep{Fougere2016b}.

As regards the stared off-limb observations considered in this paper, they probe coma regions at $PA$ close to the projected comet-Sun line. The PA of the FOV is given for each data cube in Table~\ref{tab:log}. The angle between the PA of the FOV and the main fan direction is the range 0--70$^{\circ}$ (with 70\% of the observations with an offset less than 45$^{\circ}$). Since the angular width of the H$_2$O and CO$_2$ fans was measured to be typically 110$^{\circ}$  and 180$^{\circ}$ for H$_2$O and CO$_2$, respectively \citep{dbm2016ESLAB}, coma regions within the molecular fan were observed most of the time.


\section{Observed fluorescence emissions and modelling}
\label{sec:model}

\begin{figure}
    \includegraphics[width=\columnwidth]{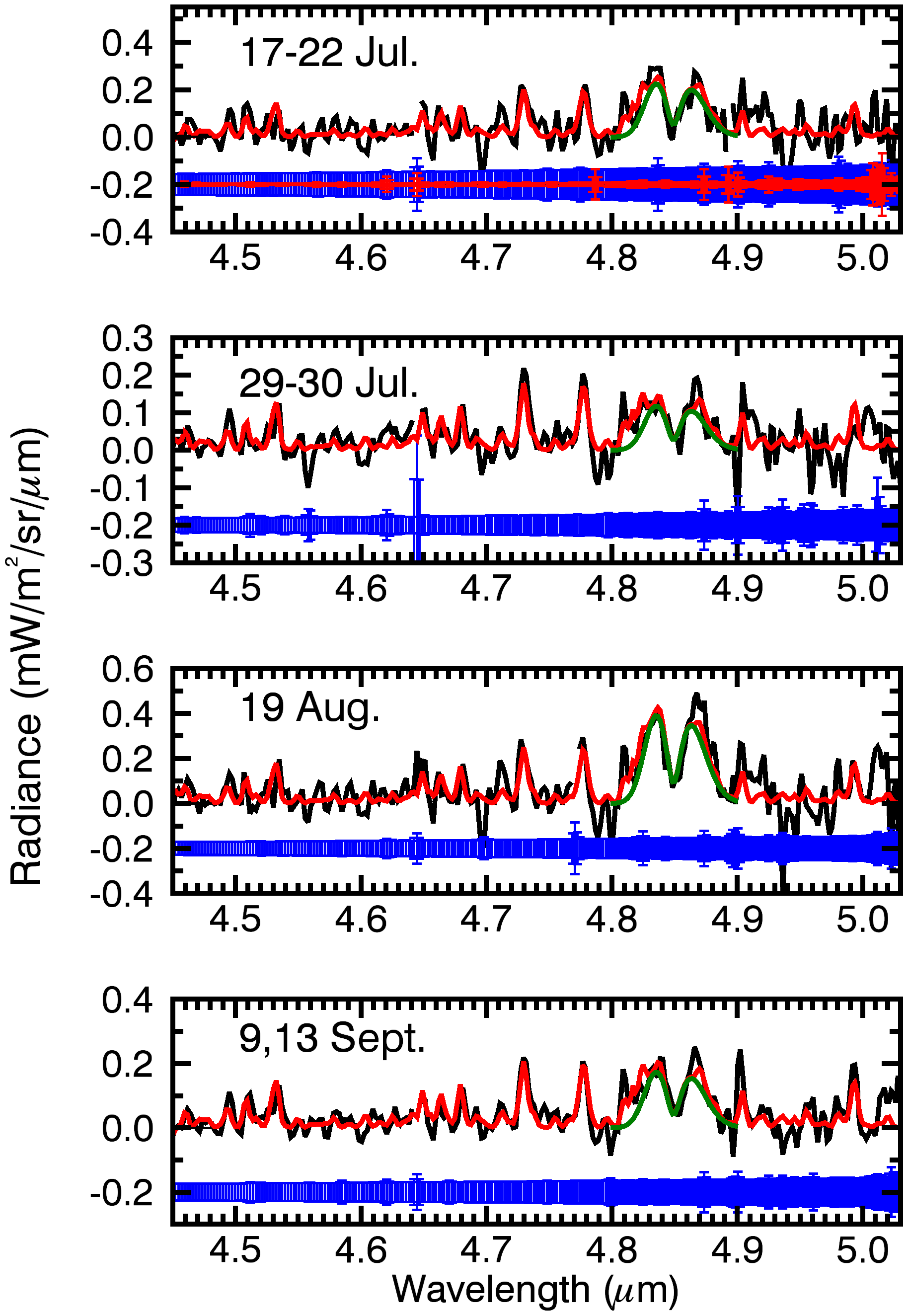}
   \caption{VIRTIS-H spectra in the range 4.45--5 $\mu$m showing ro-vibrational lines of the H$_2$O $\nu_3-\nu_2$ band, and the OCS $\nu_3$ band. The red line displays the fitted synthetic spectrum for a temperature of
   120 K, including both H$_2$O and OCS. The green line shows the contribution due to OCS only. A: 17--22 July; B : 29--30 July; C: 19 Aug.; D: 9, 13 Sept.
   The standard deviation for each spectral channel is shown in the bottom of the plots (for better reading, the errorbars are offsetted with respect to the measured radiances); values in blue include time variations in the continuum. In the top plot, the standard deviation measured on a spectrum with no comet signal (and scaled to the observing time of the comet spectrum) is superimposed, in red.}
    \label{fig:OCS-H2O}
\end{figure}

\subsection{H$_2$O}

H$_2$O is detected both in the 2.5--2.9 $\mu$m (covered by orders 3, 4 and 5 of the spectrograph) and in the 4.45--5.0 $\mu$m range (order 0) (Figs~\ref{fig:fullspectrum}-\ref{fig:OCS-H2O}).
The ro-vibrational lines detected in the 2.5--2.9 $\mu$m region are from the $\nu_3$ and $\nu_1$ fundamental bands, and from several
hot bands ($\nu_1$+$\nu_3$--$\nu_1$, $\nu_2$+$\nu_3$--$\nu_2$, and the weaker $\nu_1$+$\nu_3$--$\nu_3$ band, \citet{dbm1989,Villanueva2012}).
With the spectral resolution of the data, the ro-vibrational structure of these bands is partially resolved. The intensity of the $\nu_3$ band is expected to be significantly affected by optical depth effects \citep{Debout2016}. Unfortunately, the low wavelength range of order 3 (where the optically thin hot bands are nicely detected) is somewhat affected by straylight within the instrument, requiring careful data analysis, which is outside the scope of the present work. Therefore, we focussed on the analysis of the 4.45--5.0 $\mu$m range
where lines of the $\nu_3$--$\nu_2$ and $\nu_1$--$\nu_2$ H$_2$O hot bands are present (Fig.~\ref{fig:OCS-H2O}).

\begin{figure}
\centering
\includegraphics[width=\columnwidth]{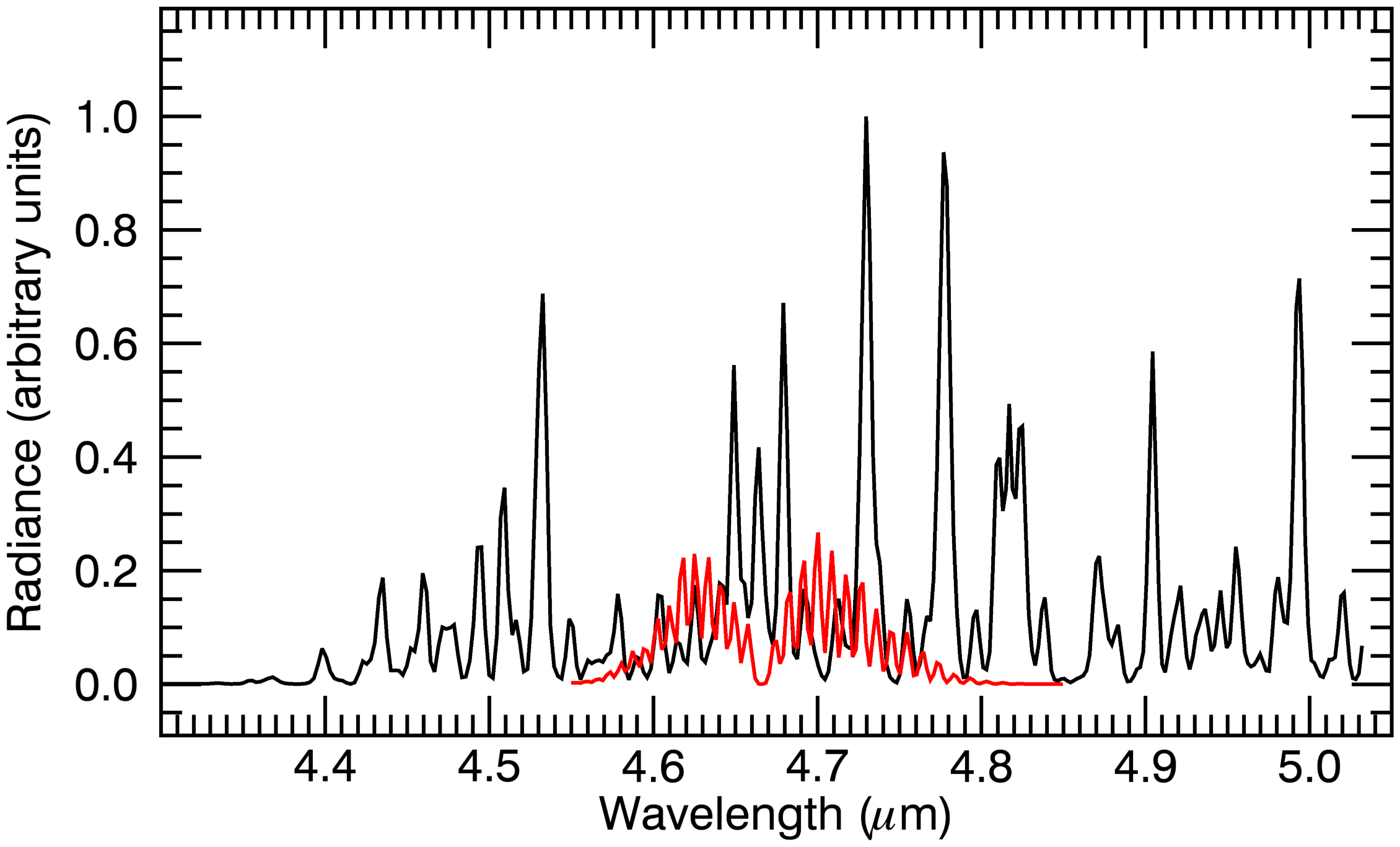}
\caption{Simulated fluorescence spectra of H$_2$O (black) and CO (red) for an effective spectral resolution of 800. The CO/H$_2$O abundance ratio is 1\% and the rotational temperature is 100 K. }
    \label{fig:mod-H2O-CO}
\end{figure}

Synthetic H$_2$O spectra were computed using the model developed by \citet{Crovisier2009}.  This model computes the full fluorescence cascade of the water molecule excited by the Sun radiation, describing the population of the rotational levels in the ground vibrational state by a Boltzmann distribution at $T_{rot}$ (the rotation temperature $T_{rot}$ is expected to be representative of the gas kinetic temperature in the inner coma of 67P). We verified that excitation by the Sun radiation scattered by the nucleus, and by the nucleus thermal radiation can be neglected (this is the case for all the molecular bands studied in the paper). The  model, which includes a number of  fundamental bands and hot bands, uses the comprehensive H$_2$O {\it ab initio} database of \citet{Schwenke2000}, and describes the solar radiation as a blackbody. Synthetic spectra in the 4.4--4.5 $\mu$m range closely resemble those obtained by the model of \citet{Villanueva2012}, which uses the BT2 {\it ab initio} database \citep{Barber2006} and includes a more exact description of the solar radiation field including solar lines. However, in that spectral region, the excitation of the water bands is not affected by solar Fraunhofer lines (Villanueva, personal communication). Finally, to achieve the best accuracy for the band excitation and emission rates, the synthetic spectra were rescaled to match those obtained by \citet{Villanueva2012}. The resulting total emission rate (g-factor) in the 4.2--5.2 $\mu$m region is given in Table~\ref{tab:g-fact} for a gas temperature of 100 K (actually, as we fitted water lines in the 4.45-5.03 $\mu$m range, we used the corresponding g-factor of 1.19 $\times$ 10$^{-5}$ s$^{-1}$ at 1 AU from Sun).
A synthetic spectrum at the spectral resolution of VIRTIS-H is shown in Fig.~\ref{fig:mod-H2O-CO}.

These models assume optically thin conditions, both in the excitation of the vibrational bands and in the received radiation. However, the observed $\nu_3$--$\nu_2$ rovibrational lines result from cascades from the $\nu_3$ vibrational state, and one may wonder whether the solar pump populating the $\nu_3$ state is attenuated in the regions probed by the LOS. Using the radiative transfer code of \citet{Debout2016}, we verified that the excitation of the  $\nu_3$ band is only weakly affected by optical depths (by 10\% at most). Because the hot bands are connecting weakly populated vibrational states, they are fully optically thin with respect to the received radiation, unlike the $\nu_3$ band at 2.67 $\mu$m.

\subsection{The CO$_2$, $^{13}$CO$_2$ complex}
\label{sec:modelCO2}

\begin{figure}
    \includegraphics[width=\columnwidth]{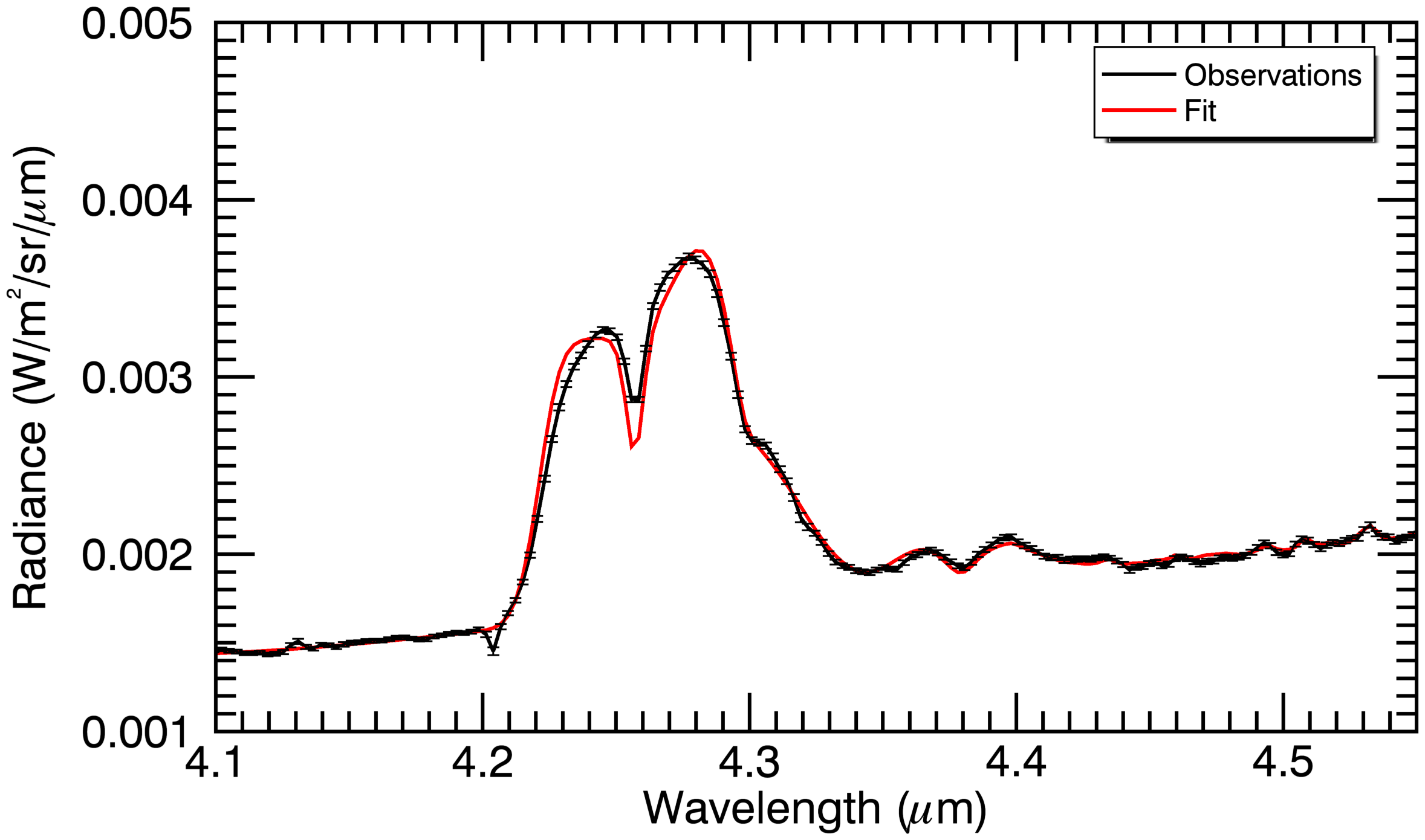}
    \includegraphics[width=\columnwidth]{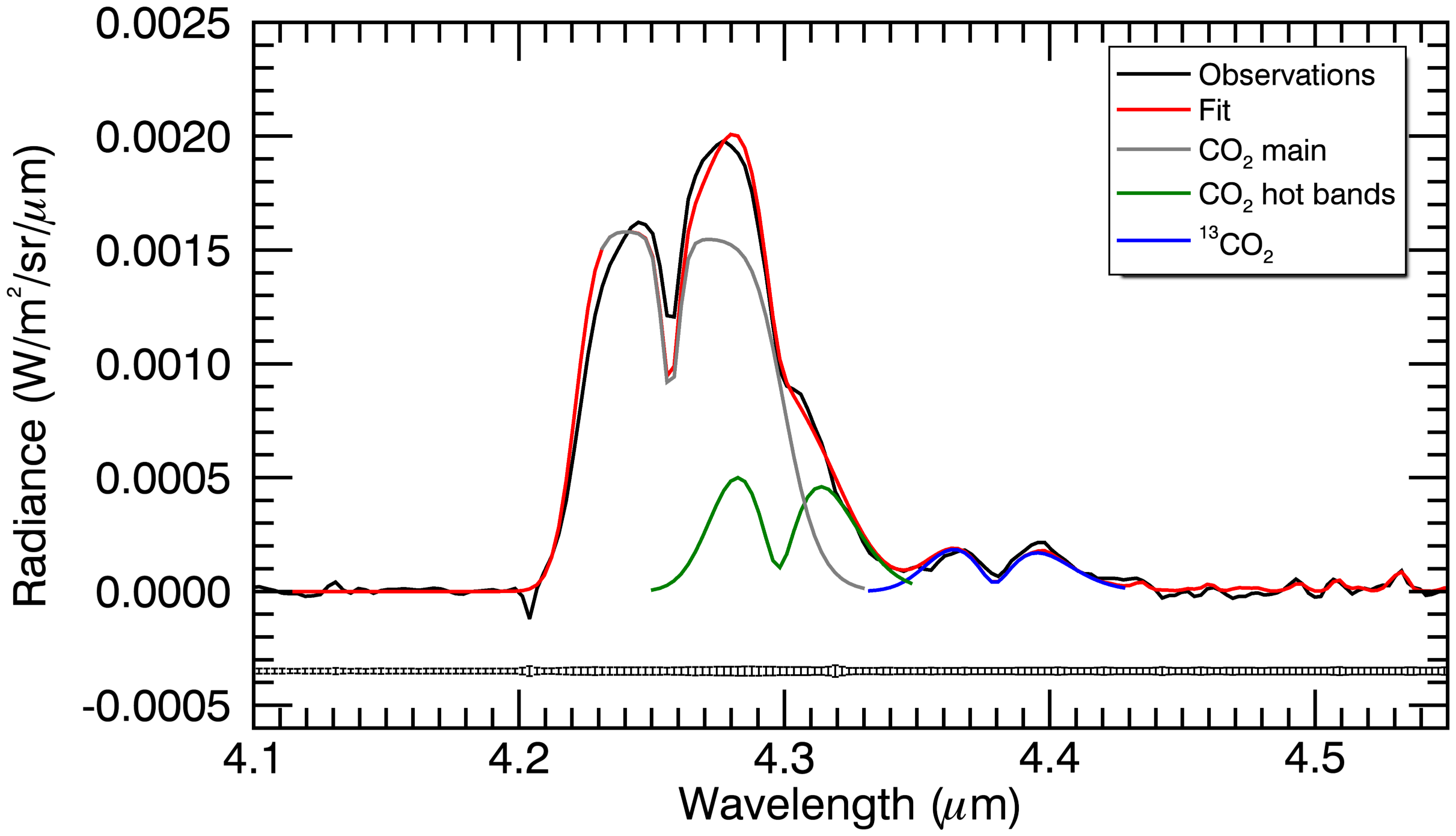}
    \caption{Exemple of model fit to a VIRTIS-H spectrum in the 4.2--4.6 $\mu$m range. The considered spectrum is the mean of four data cubes obtained on 6 and 7 September 2015.  Top panel: observed spectrum with standard deviations (black) and model fit (red). Bottom panel: observed spectrum and model fit for each individual band (grey, green and blue lines for CO$_2$, CO$_2$ hot bands, and $^{13}$CO$_2$, respectively); the continuum radiation due to dust thermal radiation has been removed; standard deviations are given in the bottom of the plot (see Fig.~\ref{fig:OCS-H2O}). Free model parameters are the baseline polynomial, and multiplying factors to CO$_2$, $^{13}$CO$_2$, and H$_2$O fluorescence spectra computed for a rotational  temperature of 120 K. The ratio between the intensities of the CO$_2$ hot bands and the $^{13}$CO$_2$ band is fixed to the expected value for $^{12}$CO/$^{13}$CO = 89. }
    \label{fig:fit-CO2}
\end{figure}

\begin{table}
    \caption{Fluorescence g-factors.}
    \label{tab:g-fact}
    \begin{tabular}{lclrl} 
        \hline
        Molecule & Band & Wavelength & g-factor$^a$ & Ref.$^b$\\
        & &$\mu$m& s$^{-1}$ & \\
        \hline
        H$_2$O & $\nu_3-\nu_2, \nu_1-\nu_2$ & 4.2--5.2 & 1.40 $\times$ 10$^{-5}$ & (1)\\
        CO$_2$ & $\nu_3$ & 4.26 & 2.69 $\times$ 10$^{-3}$ & (2) \\
        CO$_2$ & $\nu_1+\nu_3-\nu_1$ & 4.30 & 8.7 $\times$ 10$^{-5c}$ & (3) \\
        $^{13}$CO$_2$ & $\nu_3$ & 4.38 & 2.42 $\times$ 10$^{-3}$ & (4) \\
        CH$_4$ & $\nu_3$ & 3.31 & 4.0 $\times$ 10$^{-4}$ & (4) \\
        OCS & $\nu_3$  & 4.85 & 2.80 $\times$ 10$^{-3}$ & (4)\\
        \hline
    \end{tabular}
    {\footnotesize
    $^a$ The total band g-factors are listed for $r_h = 1$~AU and scale as $r_h^{-2}$. They are computed for $T_{rot} = 100$~K (but they only weakly depend upon $T_{rot}$).\\
    $^b$ References: (1) \citet{Villanueva2012}; (2) \citet{Debout2016}; (3) \citet{Crovisier2009}; (4) this work.\\
     $^c$ Sum of the g-factors of the $10^01_{II}-10^00_{II}$ and
$10^01_{I}-10^00_{I}$ bands.}
\end{table}

The CO$_2$ $\nu_3$ band shows strong fluorescence emission between
4.2 and 4.32 $\mu$m (Fig.~\ref{fig:fit-CO2}). A wing in the long
wavelength range extending up to about 4.35 $\mu$m is present. It
is attributed  to fluorescence emission of two hot bands of
CO$_2$, namely the $10^01_{II}-10^00_{II}$ and
$10^01_{I}-10^00_{I}$ ($\nu_1+\nu_3-\nu_1$ bands) centred at
2327.4 and 2326.6 cm$^{-1}$ (4.30 $\mu$m), respectively. A weak
band extending from 4.35 to 4.42 $\mu$m is present,
identified as the $^{13}$CO$_2$ $\nu_3$ band. This is the first
detection of $^{13}$CO$_2$ and CO$_2$ hot bands in a comet. In
optically thin conditions, these bands are much weaker than the
CO$_2$ main band. A priori, their observation allows a
determination of the CO$_2$ column density independently of
transfer modelling. However, some modelling of the spectral shape
of the CO$_2$ main band is required since : i) the hot bands are
blended with the $P$ branch of the CO$_2$ main band in the
observations considered here (whereas observations obtained
farther from the nucleus in the cold coma show distinctly the $R$
branch of the hot bands); ii)  the CO$_2$ complex is close to the
edge of order 0, with little margin for baseline removal.

As in the case of H$_2$O, we modelled the fluorescence excitation
of CO$_2$ and subsequent cascades in the optically thin
approximation following \citet{Crovisier2009}, using line
strengths and frequencies from the GEISA database
\citep{GEISA2011}. The g-factor for the sum of the two hot bands
is 3.3\% of that of the main band (Table~\ref{tab:g-fact}). For
the $\nu_3$ band of $^{13}$CO$_2$, synthetic spectra were computed
assuming resonant fluorescence and using the HITRAN molecular
database \citep{Rothman2013}. The blend of the two hot bands is
about three times more intense than the $^{13}$CO$_2$
$\nu_3$ band for a terrestrial $^{12}$C/$^{13}$C ratio of 89.

Fluorescence emission from the $\nu_3$ band of C$^{18}$O$^{16}$O
is not significant and will not be considered in the spectral
analysis. This band is centred at 4.29 $\mu$m, i.e., close to the
wavelength of the hot bands. Assuming the terrestrial
$^{16}$O/$^{18}$O ratio of 500, this band is estimated to be nine
times less intense than the blend of the two hot bands.

Sample spectra of the CO$_2$ $\nu_3$ band in optically thick
conditions were calculated using a simple approach. For computing
the  excitation of the excited vibrational state, we used the
Escape probability formalism, following \citet{dbm1987}. The
intensity of the individual lines [W m$^{-2}$ sr$^{-1}$] was then
computed according to:


\begin{equation}
I_{ul} = \frac{2 h c v_{th}}{\lambda^4} \frac{p_u}{\frac{w_u}{w_l}p_l-p_u} (1-e^{-\tau_{ul}}),
\end{equation}

\noindent
with the line opacity $\tau_{ul}$ computed according to:

\begin{equation}
\tau_{ul} = \frac{A_{ul} N w_u}{8 \pi \lambda^3 v_{th}} (\frac{p_l}{w_l}-\frac{p_u}{w_u}),
\end{equation}


\noindent where $\lambda$ is the wavelength, the $u$ and $l$
indices refer to the upper and lower levels of the transition,
respectively, with $p_i$, $w_i$, $A_{ij}$ corresponding to
 level populations, statistical weights,  and Einstein-$A$
coefficient, respectively. This formula assumes thermal line
widths described by the thermal velocity $v_{th}$.  $N$ is the
column density along the LOS. The input parameters of this model
are the column density and the rotational temperature. We adjusted
these parameters to obtain a CO$_2$ band shape approximately
matching the observed one. Both parameters affect the band shape,
that is the width, the relative intensity of the $P$, $R$
branches, and the depth of the valley between the two branches.
Because this model is very simplistic \citep[see exact radiative
transfer modelling of][]{Debout2016}, the input parameters that
provide good fit to the data will not be discussed.

\subsection{CH$_4$}

\begin{figure}
    \includegraphics[width=\columnwidth]{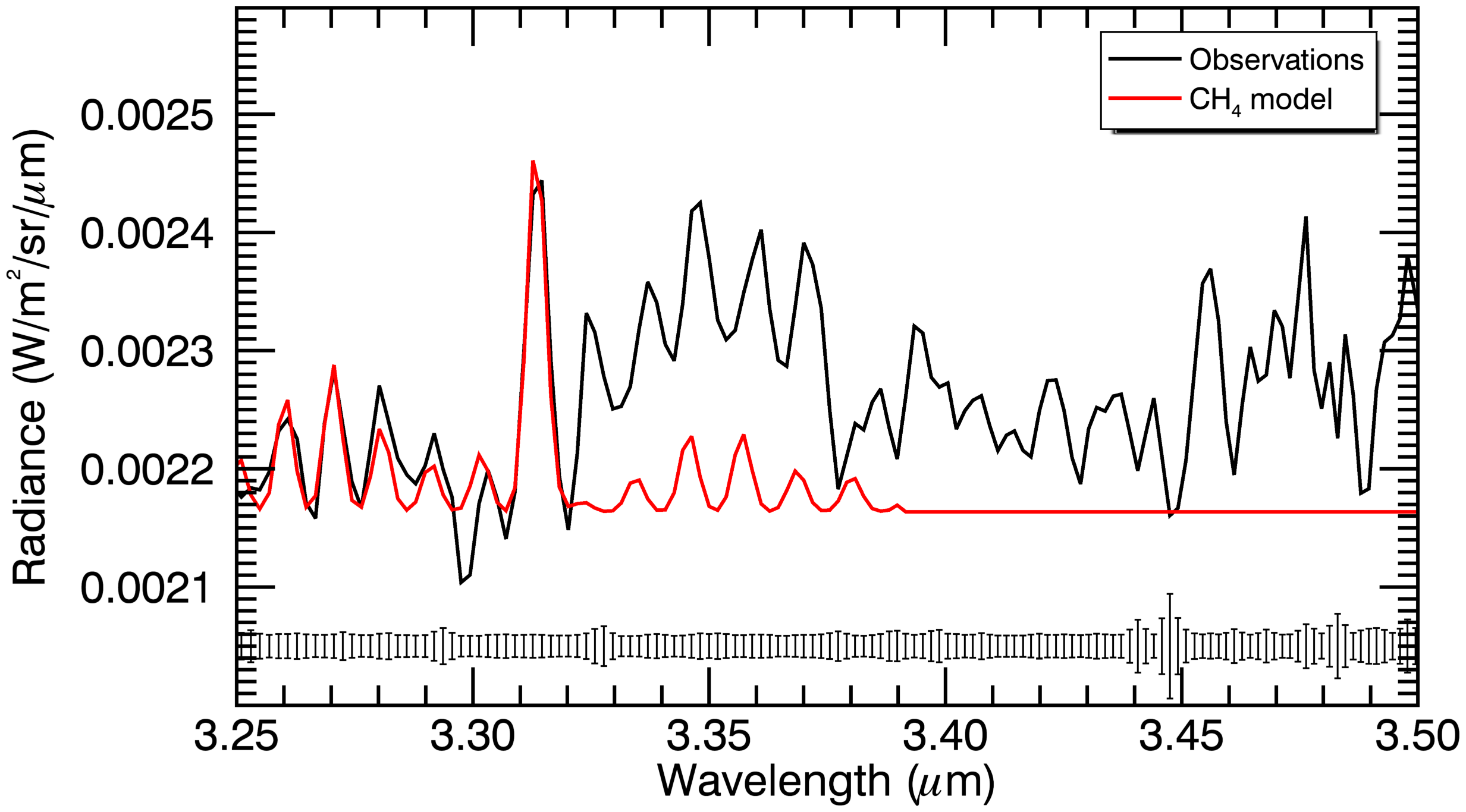}
   \caption{VIRTIS-H spectrum in order 2 covering the CH$_4$ $\nu_3$ band, obtained by averaging 9 data cubes showing well detected CH$_4$ emission (SNR > 20). The synthetic spectrum of CH$_4$ computed for a rotational  temperature of 100 K is shown in red. The standard deviations are shown in the bottom of the plot (see Fig.  \ref{fig:OCS-H2O}). Species other than CH$_4$ (in particular C$_2$H$_6$ and CH$_3$OH) contribute to the signal longward 3.32 $\mu$m. }
    \label{fig:CH4}
\end{figure}

The $\nu_3$ band of CH$_4$ at 3.31 $\mu$m presents a $P$, $Q$, $R$
structure. The strong $Q$ branch, which is undetectable from
ground-based observations due to telluric absorptions, is detected in 75\% of the considered
VIRTIS-H cubes. The $R$ lines, shortwards of 3.31 $\mu$m, are also
well detected when averaging several cubes,  with relative intensities
consistent with fluorescence modelling (Fig.~\ref{fig:CH4}).
The discrepancy between the CH$_4$ fluorescence model and
observations longward 3.32 $\mu$m seen in Fig.~\ref{fig:CH4} is
caused by the blending of CH$_4$ $P$ lines with molecular lines
from other species. Indeed, ground-based spectral surveys of this spectral
region in several comets show number of lines, especially from
C$_2$H$_6$ and CH$_3$OH \citep{Dello2006,Dello2013}. The analysis
of the 3.32--3.5 $\mu$m region, characteristic of C-H stretching
modes of organic molecules, will be the object of a forthcoming
paper.

\begin{figure}
\centering
\includegraphics[width=\columnwidth]{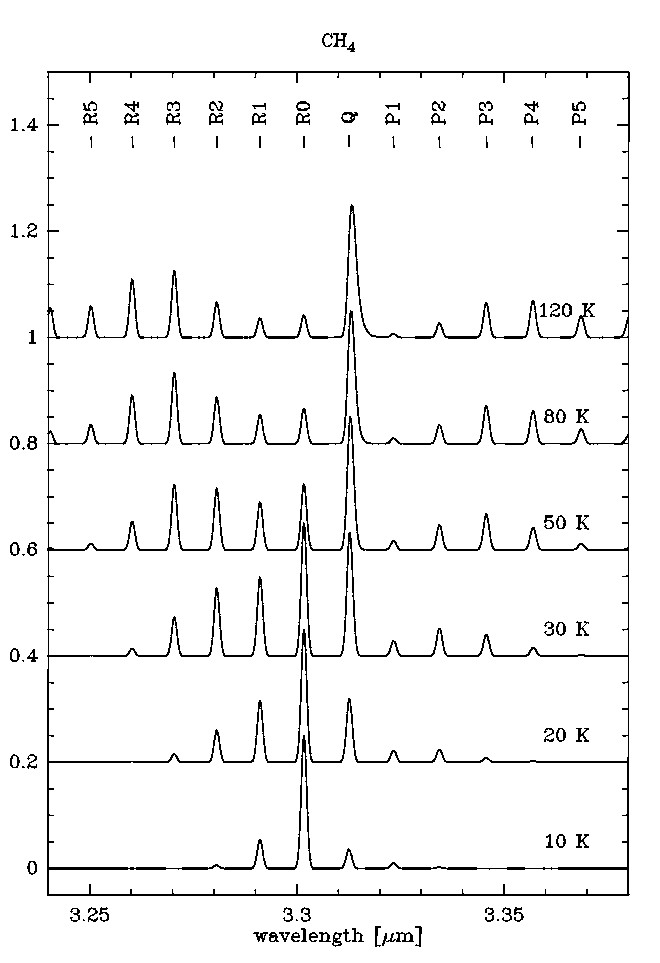}
\caption{Fluorescence spectra of methane simulated for various rotational
temperatures. The $A$, $E$ and $F$ spin species were assumed to be in
equilibrium with the rotational temperature. The spectra are
normalized to their maximum intensity; note that the spectral resolution is not adjusted to that of
the VIRTIS-H instrument.}
\label{fig:ch4_t}
\end{figure}

Synthetic spectra of the  $\nu_3$ band of CH$_4$ simulated for various temperatures are shown in Fig.~\ref{fig:ch4_t}. They were computed for resonant fluorescence using the HITRAN database \citep{Rothman2013}. Although some observations point to low nuclear spin temperatures for methane \citep{Kawakita2005}, we assumed the distribution of the $A$, $E$ and $F$ spin species to be equilibrated with the rotational temperature. The low signal-to-noise ratio does not allow us to study the spin distribution from the VIRTIS-H observations. The $Q$ branch, which is an  unresolved blend of $\Delta J = 0$ lines of the three spin species, has a g-factor which is insensitive to the rotation (and spin) temperature for $T \gtrsim 60$~K. It amounts to 30\% of the total band intensity. This branch alone will be used to estimate the CH$_4$ column density.

\subsection{OCS and CO}
\label{sec:model3}

The OCS $\nu_3$ band at 4.85 $\mu$m (C--O stretching mode) has a
strength comparable to that of the CO$_2$ $\nu_3$ band, allowing
its detection even for small column densities in infrared cometary
spectra \citep{Dello1998}. In comet 67P, it is detected in most of
the data cubes as a faint signal with a peak radiance similar to
the strongest water lines nearby (Fig.~\ref{fig:OCS-H2O}). Our
synthetic spectrum follows the model of \citet{Crovisier1987},
with a total g-factor of 2.8 $\times$ 10$^{-3}$ s$^{-1}$ ($r_h$ =
1 AU) updated from the HITRAN database \citep{Rothman2013}.

In the same spectral region, with the actual VIRTIS-H spectral
resolution, the ro-vibrational lines of the CO ($v=1-0$) band at
4.67 $\mu$m are blended with the lines of water
(Fig~\ref{fig:mod-H2O-CO}). CO has been detected in comet 67P with
the Alice and MIRO instruments (Feldman, personal communication;
Biver et al. in preparation) with an abundance relative to water
on the order of 1\%. With a band g-factor of 2.51 $\times$
10$^{-4}$ s$^{-1}$ ($r_h$ = 1 AU), based on fluorescence
calculations, CO emission is then expected to be faint in VIRTIS
spectra. Therefore, the contribution of this band to the spectrum
has been neglected in the investigation of the H$_2$O lines. In
Fig~\ref{fig:mod-H2O-CO}, the CO spectrum expected for a CO/H$_2$O
abundance ratio of 1\% is superimposed on the H$_2$O spectrum.
A tentative estimation of the CO/H$_2$O abundance based on VIRTIS-H spectra is 
provided in Sect.~\ref{sec:results}.
 

\section{Column density and abundance determinations}
\label{sec:results}

\begin{table*}
    \caption{CO$_2$, $^{13}$CO$_2$ $\nu_3$ band intensities, and $^{13}$CO$_2$ column density.}
    \label{tab:RES-CO2}
    \begin{tabular}{lllll} 
        \hline
        Start time & $I$($^{13}$CO$_2$) & $I$(CO$_2$) & CO$_2$ $\nu_3$ atten.$^a$ &  $N$($^{13}$CO$_2$) \\
(UT) & (10$^{-5}$ W m$^{-2}$ sr$^{-1}$) & (10$^{-5}$ W m$^{-2}$ sr$^{-1}$) &  & (10$^{18}$ m$^{-2}$)  \\
        \hline
2015-07-08\phantom{T}21:11:00.3 & 0.90 $\pm$ 0.40$^{b,c}$  &16.20 $\pm$ 0.09$^b$ &   3$-$7 & 1.75 $\pm$ 0.77$^{b,c}$ \\
2015-07-12\phantom{T}11:46:52.6 & $-$  & $-$ &   $-$ & $-$  \\
2015-07-17\phantom{T}08:15:55.9 & 0.90 $\pm$ 0.30$^{c}$  &15.40 $\pm$ 0.02 &   3$-$7 & 1.70 $\pm$ 0.57$^{c}$ \\
2015-07-22\phantom{T}15:23:57.9 & 1.20 $\pm$ 0.20$^{c}$  &30.90 $\pm$ 0.04 &   3.4 & 2.20 $\pm$ 0.37$^{c}$ \\
2015-07-22\phantom{T}21:10:24.7 & 0.43 $\pm$ 0.11$^{c}$  &16.50 $\pm$ 0.03 &   2.3 & 0.80 $\pm$ 0.20$^{c}$ \\
2015-07-27\phantom{T}23:06:40.5 & 0.53 $\pm$ 0.09$^{c}$  &14.00 $\pm$ 0.01 &   2.8 & 0.97 $\pm$ 0.16$^{c}$ \\
2015-07-29\phantom{T}21:20:59.8 & 0.53 $\pm$ 0.03  &17.90 $\pm$ 0.04 &   2.6 & 0.95 $\pm$ 0.06 \\
2015-07-30\phantom{T}01:47:25.8 & 0.43 $\pm$ 0.03  &17.70 $\pm$ 0.04 &   2.2 & 0.78 $\pm$ 0.06 \\
2015-08-01\phantom{T}08:25:16.8 & 0.87 $\pm$ 0.03  &22.70 $\pm$ 0.04 &   3.4 & 1.57 $\pm$ 0.05 \\
2015-08-10\phantom{T}23:32:29.0 & 0.35 $\pm$ 0.03  &12.80 $\pm$ 0.04 &   2.4 & 0.62 $\pm$ 0.06 \\
2015-08-16\phantom{T}11:56:25.1 & 1.27 $\pm$ 0.05  &23.20 $\pm$ 0.09 &   4.9 & 2.25 $\pm$ 0.09 \\
2015-08-19\phantom{T}11:08:06.5 & 2.34 $\pm$ 0.03  &32.50 $\pm$ 0.05 &   6.4 & 4.18 $\pm$ 0.06 \\
2015-08-19\phantom{T}15:34:33.4 & 2.03 $\pm$ 0.04  &31.30 $\pm$ 0.07 &   5.8 & 3.62 $\pm$ 0.08 \\
2015-08-19\phantom{T}21:20:58.5 & 2.07 $\pm$ 0.05  &24.50 $\pm$ 0.07 &   7.5 & 3.69 $\pm$ 0.08 \\
2015-08-20\phantom{T}22:13:06.6 & 1.51 $\pm$ 0.03  &16.80 $\pm$ 0.04 &   8.0 & 2.69 $\pm$ 0.05 \\
2015-08-23\phantom{T}17:07:54.5 & 1.63 $\pm$ 0.03  &21.40 $\pm$ 0.05 &   6.8 & 2.92 $\pm$ 0.05 \\
2015-08-23\phantom{T}21:29:29.5 & 1.73 $\pm$ 0.03  &20.20 $\pm$ 0.04 &   7.6 & 3.10 $\pm$ 0.05 \\
2015-08-26\phantom{T}08:17:52.6 & 1.59 $\pm$ 0.03  &18.70 $\pm$ 0.04 &   7.6 & 2.86 $\pm$ 0.06 \\
2015-08-26\phantom{T}10:58:13.4 & 1.90 $\pm$ 0.08  &19.30 $\pm$ 0.11 &   8.8 & 3.44 $\pm$ 0.14 \\
2015-09-06\phantom{T}21:25:16.9 & 0.91 $\pm$ 0.04  &12.70 $\pm$ 0.07 &   6.4 & 1.71 $\pm$ 0.07 \\
2015-09-07\phantom{T}03:06:16.9 & 0.96 $\pm$ 0.04  &11.60 $\pm$ 0.07 &   7.4 & 1.81 $\pm$ 0.07 \\
2015-09-07\phantom{T}07:28:16.8 & 0.90 $\pm$ 0.05  &11.00 $\pm$ 0.07 &   7.3 & 1.70 $\pm$ 0.09 \\
2015-09-07\phantom{T}13:09:16.0 & 0.94 $\pm$ 0.03  &12.60 $\pm$ 0.05 &   6.7 & 1.78 $\pm$ 0.05 \\
2015-09-09\phantom{T}08:15:16.0 & 2.03 $\pm$ 0.03  &20.90 $\pm$ 0.04 &   8.7 & 3.87 $\pm$ 0.06 \\
2015-09-09\phantom{T}15:29:34.0 & 1.22 $\pm$ 0.02  &15.40 $\pm$ 0.03 &   7.0 & 2.32 $\pm$ 0.04 \\
2015-09-10\phantom{T}01:37:26.0 & 0.90 $\pm$ 0.02  &11.10 $\pm$ 0.02 &   7.2 & 1.71 $\pm$ 0.03 \\
2015-09-13\phantom{T}06:54:44.2 & 1.44 $\pm$ 0.02  &16.90 $\pm$ 0.02 &   7.6 & 2.81 $\pm$ 0.04 \\
2015-09-13\phantom{T}11:46:19.3 & 1.77 $\pm$ 0.08  &18.50 $\pm$ 0.13 &   8.5 & 3.45 $\pm$ 0.16 \\
2015-09-24\phantom{T}11:55:21.0 & 0.64 $\pm$ 0.01  & 9.06 $\pm$ 0.02 &   6.3 & 1.34 $\pm$ 0.03 \\
2015-09-27\phantom{T}16:57:54.5 & 0.32 $\pm$ 0.01  & 4.89 $\pm$ 0.01 &   5.8 & 0.68 $\pm$ 0.03 \\
        \hline
    \end{tabular}

    {\raggedright
$^a$  Attenuation factor of the CO$_2$ $\nu_3$ band, corresponding to the ratio between the $\nu_3$ band intensity expected in optically thin conditions and the observed $\nu_3$ band intensity (see text).   
    
$^b$ Results obtained by averaging the data of 8 and 12 July.

$^c$ Values and uncertainties  cover results obtained using either multi-component fit or the $^{13}$CO$_2$ band area measured on the spectrum.

  }
\end{table*}

\begin{table*}
    \caption{H$_2$O, CH$_4$ and OCS band intensities and column densities.}
    \label{tab:RES-CH4}
    \begin{tabular}{lcccccc} 
        \hline
        Start time & \multicolumn{2}{c}{H$_2$O} & \multicolumn{2}{c}{CH$_4$} & \multicolumn{2}{c}{OCS}\\
                   & \multicolumn{2}{c}{-------------------------------------} & \multicolumn{2}{c}{-------------------------------------} & \multicolumn{2}{c}{-------------------------------------}\\
         & $I$ &  $N^b$ & $I^a$&  $N^c$& $I^a$&  $N^c$ \\
(UT) &  (10$^{-5}$ W m$^{-2}$ sr$^{-1}$) & (10$^{20}$ m$^{-2}$) & (10$^{-6}$ W m$^{-2}$ sr$^{-1}$) & (10$^{18}$ m$^{-2}$) & (10$^{-6}$ W m$^{-2}$ sr$^{-1}$) & (10$^{18}$ m$^{-2}$)   \\
        \hline
2015-07-08\phantom{T}21:11:00.3 & 1.82 $\pm$ 0.56  & 9.25 $\pm$ 2.88 & $-$\phantom{0.69 $\pm$ 0.48}  &$-$\phantom{0.62 $\pm$ 0.44} &  6.87 $\pm$ 3.34 & 1.31 $\pm$ 0.63 \\
2015-07-12\phantom{T}11:46:52.6 & 1.90 $\pm$ 0.14  &10.10 $\pm$ 0.77 & $-$\phantom{3.42 $\pm$ 0.17}  & $-$\phantom{3.04 $\pm$ 0.15} &  $-$\phantom{2.63 $\pm$ 0.86} & $-$\phantom{0.49 $\pm$ 0.16} \\
2015-07-17\phantom{T}08:15:55.9 & 1.22 $\pm$ 0.07  & 6.33 $\pm$ 0.36 & $-$\phantom{0.56 $\pm$ 0.14}  &$-$\phantom{0.49 $\pm$ 0.12} &  8.82 $\pm$ 0.40 & 1.60 $\pm$ 0.07 \\
2015-07-22\phantom{T}15:23:57.9 & 1.86 $\pm$ 0.08  &10.00 $\pm$ 0.41 & 3.50 $\pm$ 0.13  &2.97 $\pm$ 0.11 &  9.50 $\pm$ 0.48 & 1.68 $\pm$ 0.09 \\
2015-07-22\phantom{T}21:10:24.7 & 1.90 $\pm$ 0.05  & 9.79 $\pm$ 0.28 & $-$\phantom{1.84 $\pm$ 0.12}  &$-$\phantom{1.56 $\pm$ 0.10} &  3.49 $\pm$ 0.37 & 0.62 $\pm$ 0.06 \\
2015-07-27\phantom{T}23:06:40.5 & 1.49 $\pm$ 0.04  & 6.48 $\pm$ 0.17 & 1.78 $\pm$ 0.10  &1.48 $\pm$ 0.08 &  2.03 $\pm$ 0.28 & 0.35 $\pm$ 0.05 \\
2015-07-29\phantom{T}21:20:59.8 & 1.49 $\pm$ 0.10  & 6.62 $\pm$ 0.44 & $-$\phantom{1.10 $\pm$ 0.17}  &$-$\phantom{0.91 $\pm$ 0.14} &  5.00 $\pm$ 0.61 & 0.87 $\pm$ 0.11 \\
2015-07-30\phantom{T}01:47:25.8 & 1.63 $\pm$ 0.11  & 7.27 $\pm$ 0.48 & 1.85 $\pm$ 0.17  &1.53 $\pm$ 0.14 &  4.93 $\pm$ 0.64 & 0.85 $\pm$ 0.11 \\
2015-08-01\phantom{T}08:25:16.8 & 1.63 $\pm$ 0.09  & 7.10 $\pm$ 0.39 & 2.99 $\pm$ 0.35  &2.46 $\pm$ 0.29 &  5.93 $\pm$ 0.63 & 1.02 $\pm$ 0.11 \\
2015-08-10\phantom{T}23:32:29.0 & 1.17 $\pm$ 0.11  & 4.19 $\pm$ 0.40 & 1.72 $\pm$ 0.29  &1.40 $\pm$ 0.24 &  $-$\phantom{2.99 $\pm$ 0.67} & $-$\phantom{0.51 $\pm$ 0.11} \\
2015-08-16\phantom{T}11:56:25.1 & 2.01 $\pm$ 0.14  & 9.21 $\pm$ 0.63 & 2.58 $\pm$ 0.25  &2.09 $\pm$ 0.20 &  6.43 $\pm$ 0.86 & 1.09 $\pm$ 0.15 \\
2015-08-19\phantom{T}11:08:06.5 & 2.02 $\pm$ 0.10  &10.20 $\pm$ 0.50 & 7.77 $\pm$ 0.16  &6.33 $\pm$ 0.13 & 16.50 $\pm$ 0.65 & 2.82 $\pm$ 0.11 \\
2015-08-19\phantom{T}15:34:33.4 & 2.42 $\pm$ 0.13  &11.60 $\pm$ 0.64 & 5.40 $\pm$ 0.18  &4.40 $\pm$ 0.15 & 16.90 $\pm$ 0.79 & 2.89 $\pm$ 0.13 \\
2015-08-19\phantom{T}21:20:58.5 & 1.97 $\pm$ 0.13  & 8.90 $\pm$ 0.61 & 4.93 $\pm$ 0.22  &4.02 $\pm$ 0.18 & 13.80 $\pm$ 0.79 & 2.36 $\pm$ 0.14 \\
2015-08-20\phantom{T}22:13:06.6 & 2.03 $\pm$ 0.09  & 8.32 $\pm$ 0.38 & 6.83 $\pm$ 0.17  &5.57 $\pm$ 0.14 &  9.54 $\pm$ 0.53 & 1.63 $\pm$ 0.09 \\
2015-08-23\phantom{T}17:07:54.5 & 1.89 $\pm$ 0.08  & 8.27 $\pm$ 0.33 & 4.52 $\pm$ 0.18  &3.71 $\pm$ 0.15 &  9.03 $\pm$ 0.48 & 1.55 $\pm$ 0.08 \\
2015-08-23\phantom{T}21:29:29.5 & 1.65 $\pm$ 0.08  & 7.44 $\pm$ 0.36 & 5.53 $\pm$ 0.24  &4.54 $\pm$ 0.20 &  8.07 $\pm$ 0.51 & 1.39 $\pm$ 0.09 \\
2015-08-26\phantom{T}08:17:52.6 & 1.63 $\pm$ 0.08  & 6.87 $\pm$ 0.36 & 4.92 $\pm$ 0.41  &4.06 $\pm$ 0.34 &  6.01 $\pm$ 0.70 & 1.04 $\pm$ 0.12 \\
2015-08-26\phantom{T}10:58:13.4 & 2.14 $\pm$ 0.23  & 9.30 $\pm$ 0.99 & 7.38 $\pm$ 0.36  &6.09 $\pm$ 0.30 &  7.73 $\pm$ 1.30 & 1.34 $\pm$ 0.23 \\
2015-09-06\phantom{T}21:25:16.9 & 1.28 $\pm$ 0.12  & 5.25 $\pm$ 0.49 & 2.04 $\pm$ 0.22  &1.76 $\pm$ 0.19 &  $-$\phantom{2.11 $\pm$ 0.69} & $-$\phantom{0.38 $\pm$ 0.12} \\
2015-09-07\phantom{T}03:06:16.9 & 1.37 $\pm$ 0.11  & 5.64 $\pm$ 0.43 & $-$\phantom{0.99 $\pm$ 0.20}  &$-$\phantom{0.85 $\pm$ 0.17} &  3.91 $\pm$ 0.64 & 0.71 $\pm$ 0.12 \\
2015-09-07\phantom{T}07:28:16.8 & 0.89 $\pm$ 0.15  & 3.64 $\pm$ 0.60 & 2.23 $\pm$ 0.25  &1.93 $\pm$ 0.22 &  $-$\phantom{1.98 $\pm$ 0.83} & $-$\phantom{0.36 $\pm$ 0.15} \\
2015-09-07\phantom{T}13:09:16.0 & 1.30 $\pm$ 0.09  & 5.35 $\pm$ 0.35 & 3.71 $\pm$ 0.19  &3.20 $\pm$ 0.16 &  $-$\phantom{3.32 $\pm$ 0.54} & $-$\phantom{0.60 $\pm$ 0.10} \\
2015-09-09\phantom{T}08:15:16.0 & 1.60 $\pm$ 0.09  & 8.22 $\pm$ 0.49 & 3.33 $\pm$ 0.24  &2.90 $\pm$ 0.21 &  8.76 $\pm$ 0.59 & 1.59 $\pm$ 0.11 \\
2015-09-09\phantom{T}15:29:34.0 & 1.56 $\pm$ 0.06  & 7.42 $\pm$ 0.30 & 4.12 $\pm$ 0.12  &3.59 $\pm$ 0.11 &  6.13 $\pm$ 0.41 & 1.12 $\pm$ 0.08 \\
2015-09-10\phantom{T}01:37:26.0 & 1.23 $\pm$ 0.04  & 4.93 $\pm$ 0.18 & $-$\phantom{3.69 $\pm$ 0.23}  &$-$\phantom{3.23 $\pm$ 0.20} &  3.56 $\pm$ 0.34 & 0.65 $\pm$ 0.06 \\
2015-09-13\phantom{T}06:54:44.2 & 1.47 $\pm$ 0.06  & 7.13 $\pm$ 0.29 & 3.71 $\pm$ 0.13  &3.30 $\pm$ 0.12 &  7.40 $\pm$ 0.39 & 1.38 $\pm$ 0.07 \\
2015-09-13\phantom{T}11:46:19.3 & 2.11 $\pm$ 0.21  &10.00 $\pm$ 1.00 & $-$\phantom{2.93 $\pm$ 0.59}  & $-$\phantom{2.61 $\pm$ 0.53} &  7.41 $\pm$ 1.13 & 1.38 $\pm$ 0.21 \\
2015-09-24\phantom{T}11:55:21.0 & 0.81 $\pm$ 0.04  & 3.68 $\pm$ 0.19 & 4.07 $\pm$ 0.13  &3.88 $\pm$ 0.12 & $-$\phantom{-0.06$\pm$ 0.32} &$-$\phantom{-0.01$\pm$ 0.06} \\
2015-09-27\phantom{T}16:57:54.5 & 0.56 $\pm$ 0.04  & 2.41 $\pm$ 0.18 & 2.15 $\pm$ 0.13  &2.10 $\pm$ 0.12 &  $-$\phantom{0.87 $\pm$ 0.30} & $-$\phantom{0.18 $\pm$ 0.06} \\
        \hline
    \end{tabular}

        {\raggedright

$^a$ Only values obtained when detection are given.

$^b$ Derived assuming a gas kinetic temperature varying with distance to comet centre ($\rho$) following $T_{kin}$ = (4/$\rho$[km]) $\times$ 100 K (with $T_{rot}$ = $T_{kin}$) and a g-factor of 1.19 $\times$ 10$^{-5}$ at 1 AU from the Sun pertaining to the 4.45--5.03 $\mu$m range.

$^c$ Derived assuming $T_{rot}$ = 100~K in the fluorescence model.

}
\end{table*}

\subsection{Fitting procedure}
Observed spectra were analysed using the Levenberg-Marquardt $\chi^2$ minimization algorithm, describing the spectra as a linear combination of a background continuum (polynomial) and fluorescence molecular emissions, modelled as presented in Sect.~\ref{sec:model} and convolved to the spectral resolution of the analysed data (Sect.~\ref{sec:VH}).

In the fitting procedure, the rotational temperature of the molecules was taken as a fixed parameter. In dense parts of the coma, the rotational states in the ground-vibrational state of the molecules are expected to be thermalized to the gas kinetic temperature, so that $T_{rot}$ = $T_{kin}$ locally. However, since the coma is not isothermal, VIRTIS spectra mix contributions of molecules at various $T_{rot}$, implying a ro-vibrational structure which depends in a complex way on the temperature and gas distributions, but which should reflect conditions in the denser parts of the coma probed by the VIRTIS LOS. We adopted a simplistic approach, and fixed $T_{rot}$ according to available information for comet 67P. Indeed, molecular data obtained by the MIRO instrument onboard Rosetta show that the gas kinetic temperature follows in first order $T_{kin}$[K] = (4 [km]/$\rho$[km]) $\times$ 100 [K]  for $\rho$ in the range 3--25 km, where $\rho$ is the distance to comet centre (Biver, personal communication). The corresponding temperatures we used for the VIRTIS-H data showing detected CH$_4$ and OCS emissions are thus  80--130 K. We found that adopting specific temperatures for each data cube does not change significantly the results for CH$_4$ and OCS since the intensities of the CH$_4$ $Q$ branch and of the OCS band are not sensitive to the temperature (above 60 K for CH$_4$). We assumed $T_{rot}$ = 100 K for the fluorescence emissions of CH$_4$ and OCS. For H$_2$O, CO$_2$ hot-bands, and $^{13}$CO$_2$, the fit was performed for appropriate values of $T_{kin}$.

We used optically thick CO$_2$ spectra (Sect.~\ref{sec:modelCO2})
that provide reasonable fits to the CO$_2$ $\nu_3$ band and to the
overall CO$_2$ complex, with a multiplying coefficient to the
synthetic spectrum as a free parameter. The
$^{12}$CO$_2$/$^{13}$CO$_2$ relative abundance was assumed to be
terrestrial, that is the intensity of the CO$_2$ hot-bands
relative to the $^{13}$CO$_2$ band was fixed to be equal to 89
times the ratio of the g-factors.

The CO$_2$ complex was analysed by fitting the whole 4.05--4.8
$\mu$m region, hence excluding the OCS band but including
H$_2$O emission lines. Conversely, the H$_2$O and OCS emissions
were studied considering the 4.45--5.01 $\mu$m region, i.e.,
excluding the CO$_2$ complex. CO fluorescence emission was assumed
to be negligible (Sect~\ref{sec:model3}). For the CH$_4$ band,
the fit was performed in the 3.25--3.322 $\mu$m region, ignoring 
the region where $P$ lines are blended with other
molecular lines. The fitting algorithm was successful 
in fitting the CH$_4$ $Q$ branch. The degree of the polynomial
used to fit the baseline was fixed to 2 for the CH$_4$ study, to 4
for H$_2$O and OCS, and to 5 for the CO$_2$ complex. Smaller
polynomial degrees provided often unsatisfactory results for the
continuum background observed in order 0, which is due to dust
thermal emission superimposed with some instrumental stray light
at the shortest wavelengths (see Sect.~\ref{sec:VH}).

\begin{figure}
    \includegraphics[width=\columnwidth]{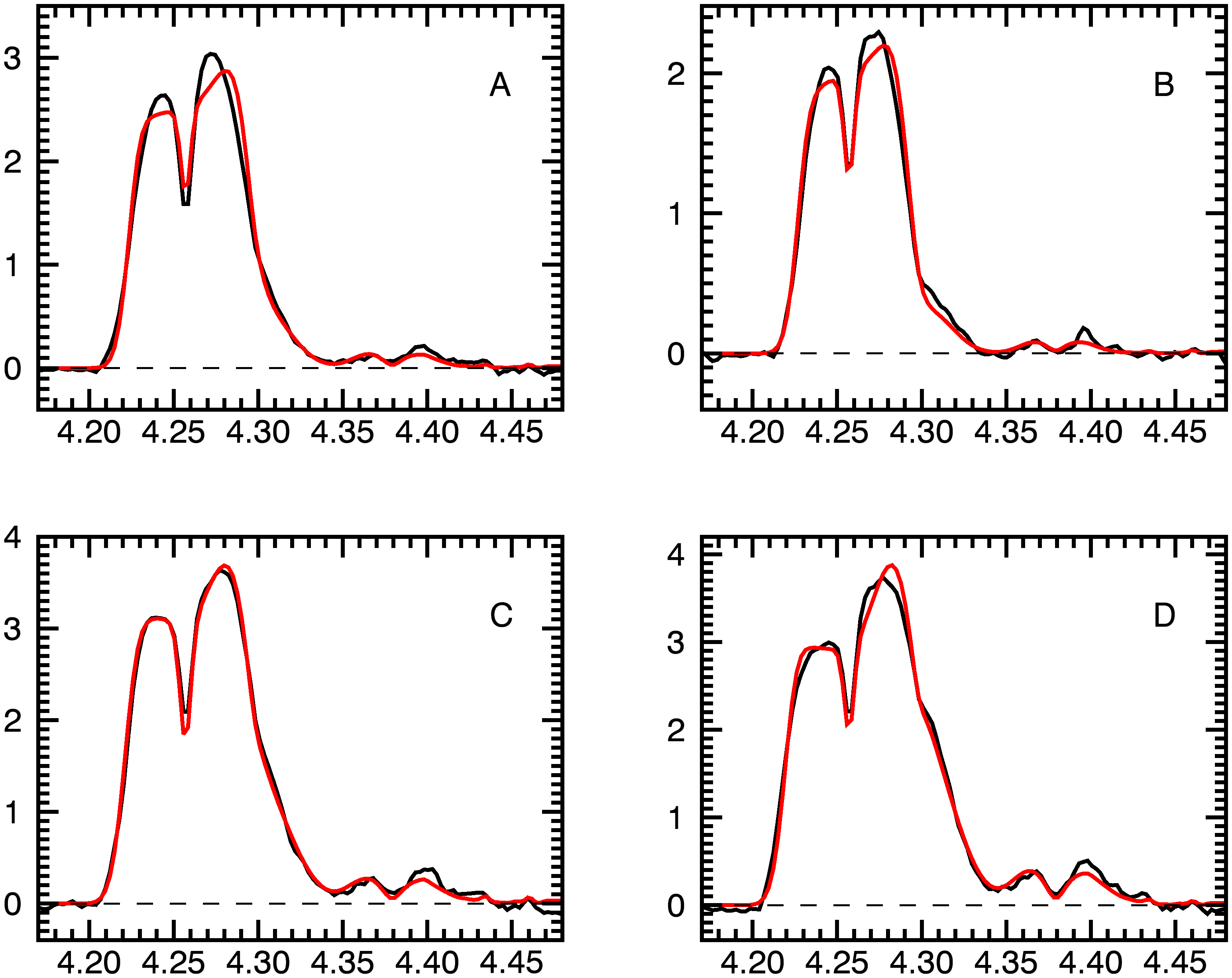}
    \vspace{0.3cm} \\
        \includegraphics[width=\columnwidth]{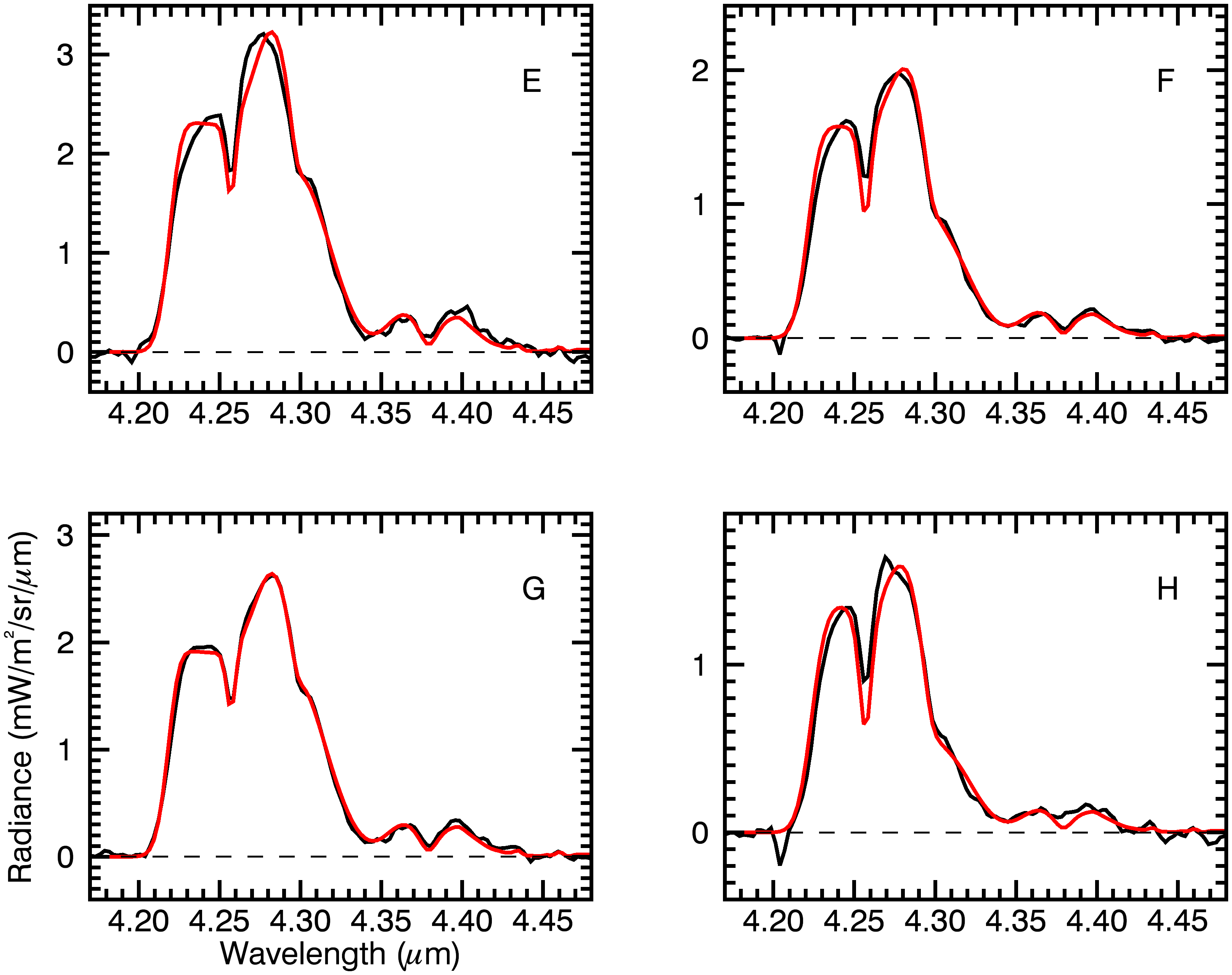}
    \caption{VIRTIS-H spectra in the 4.2--4.5 $\mu$m region (black), and model fits (red) to the CO$_2$ and $^{13}$CO$_2$ bands. The continuum background is removed. {\bf A} : 29--30 Jul., 1 Aug.; {\bf B}: 10 Aug.; {\bf C}: 16 Aug.; {\bf D}: 19--23 Aug; {\bf E}: 26 Aug.; {\bf F}: 6--7 Sept; {\bf G}: 9--13 Sept.; {\bf H}: 24 Sept.  }
    \label{fig:CO2-multi}
\end{figure}

An example of the result of the fitting procedure to the CO$_2$
complex is presented in Fig.~\ref{fig:fit-CO2}, where the
individual components are plotted. Other fits for different data sets are illustrated
in Fig.~\ref{fig:CO2-multi}. For most
individual data cubes, the $^{13}$CO$_2$ band is detected 
distinctly and the band intensity retrieved from the fit is well
estimated. However, for the six first data cubes (8--27 July), the
method led to an underestimation of the $^{13}$CO$_2$ signal (as
observed visually): the model fit fell below the observed
$^{13}$CO$_2$ emission. For these 6 observations, we also computed
the intensity in the 4.34--4.42  $\mu$m range lying above the
background, and the intensities computed by the two methods were
taken as providing the range of possible values for the
$^{13}$CO$_2$ band intensities. We note that often the model fails
to fully reproduce the intensity of the $P$ branch of
$^{13}$CO$_2$ (some residual is present at 4.4 $\mu$m,
Fig.~\ref{fig:CO2-multi}). The origin of this residual is unclear.

Examples of model fits to the  H$_2$O and OCS emissions are shown
in Fig.~\ref{fig:OCS-H2O}. For this spectral region, we identified
a shift by about 1 spectral channel (2.2 nm) between the expected
and observed position of the water lines. The shift is observed
specifically for the lines in the 4.6--4.8 $\mu$m range, whereas
the water lines shortward 4.6 $\mu$m suggest an error in the
present spectral calibration by 1/2 channel. Therefore, the
observed spectra were shifted by one channel when fitting the
H$_2$O and OCS emissions (also for Fig.~\ref{fig:OCS-H2O}).
However, we didn't apply any correction for fitting the CO$_2$
complex. Earth and Mars spectra taken with VIRTIS-H
suggests an error of 1/2 pixel in the spectral registration. We
expect further progress in the VIRTIS-H spectral calibration in the future.
A close examination of the 67P spectra plotted in Fig.~\ref{fig:OCS-H2O} shows some mismatch for the intensity of the H$_2$O lines, that may deserves further studies. In particular, the line at 4.81 $\mu$m is observed in all 67P spectra, and is most likely of cometary origin. Though water emission is expected at this wavelength, the intensity of this line is not consistent with water synthetic spectra (Figs~\ref{fig:OCS-H2O}-- \ref{fig:mod-H2O-CO}).

A critical point in the analysis is the estimation of
uncertainties. For the staring observations, the root-mean-square (rms) deviations can be estimated
for each spectral channel by studying the variation with time of
their intensity (computation of variance, then rms). However, for
data cubes showing comet signal, the rms computed in this way are larger than
statistical instrumental noise, since part of the variation is
intrinsic to the comet and/or pointing mode. All uncertainties
provided in the tables and figures include the dispersion due
to the varying intrinsic comet signal (including the continuum). In
Fig.~\ref{fig:OCS-H2O}A, we have superimposed the rms measured on
a data cube presenting little comet signal, which is effectively
much lower than the one computed with the adopted  method.
However, based on the fluctuations seen on the spectra, we believe
that using instead these rms would underestimate uncertainties.

A number of spectral channels are much noiser than their neighbors
(so-called "hot pixels"). "Hot pixels" can be spotted in
Figs~\ref{fig:OCS-H2O}, \ref{fig:fit-CO2}, and \ref{fig:CH4} from
their higher errors. There are no hot pixels for the channels covering the CH$_4$ $Q$-branch 
(Fig.~\ref{fig:CH4}). The fitting of H$_2$O lines was
performed by ignoring the very few "hot pixels" near the water
lines. The OCS band is also affected by hot pixels (see
Fig.~\ref{fig:OCS-H2O}). These were not masked in the fitting
analysis. The discrepancy observed at 4.87 $\mu$m between model
fit and observed OCS band (see Fig.~\ref{fig:OCS-H2O}) is possibly
of instrumental origin.



\subsection{Results}

\begin{figure}
    \includegraphics[width=8cm]{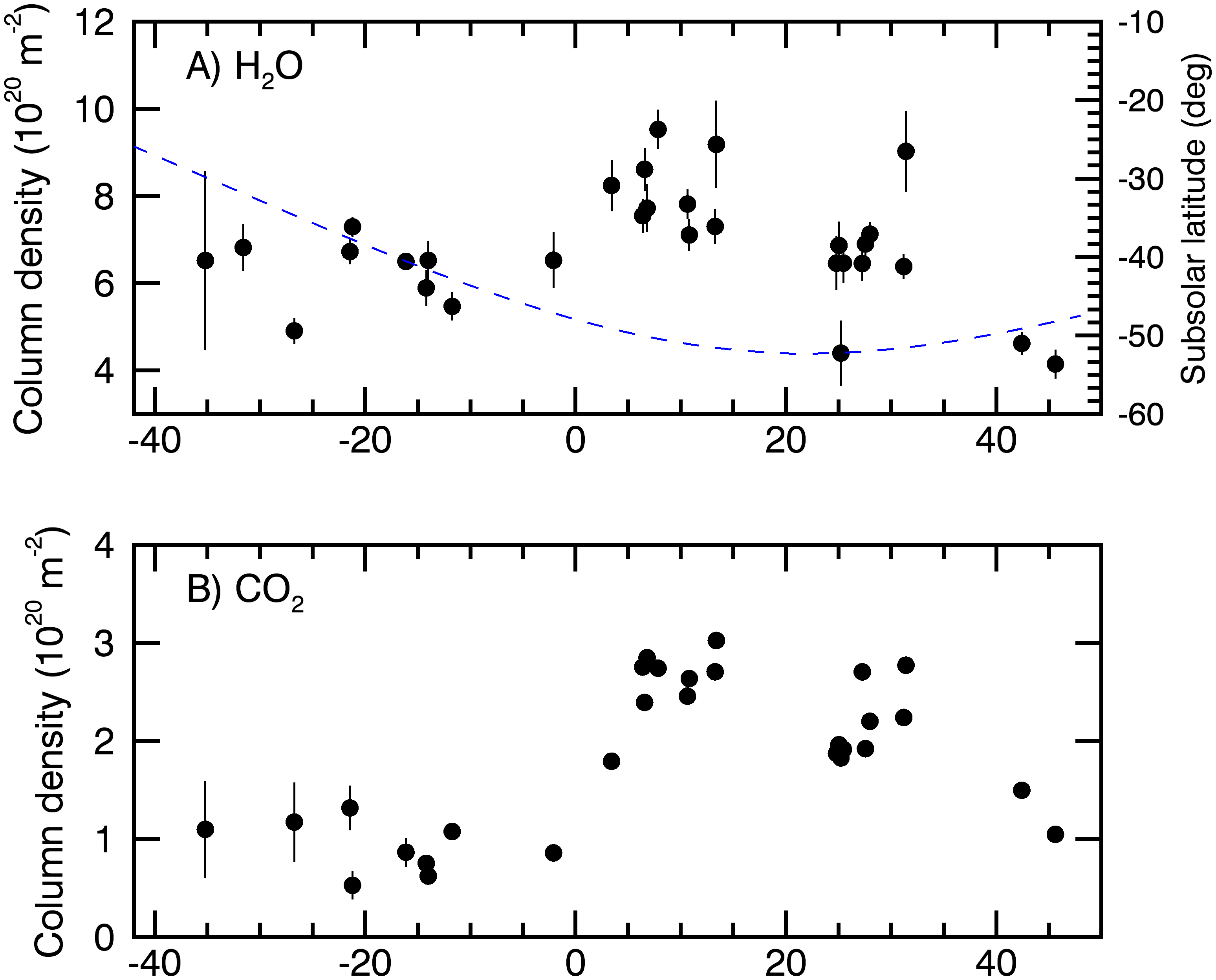}
    \par
    \vspace{0.4cm}
    \hspace{0.cm}
    \includegraphics[width=7.23cm]{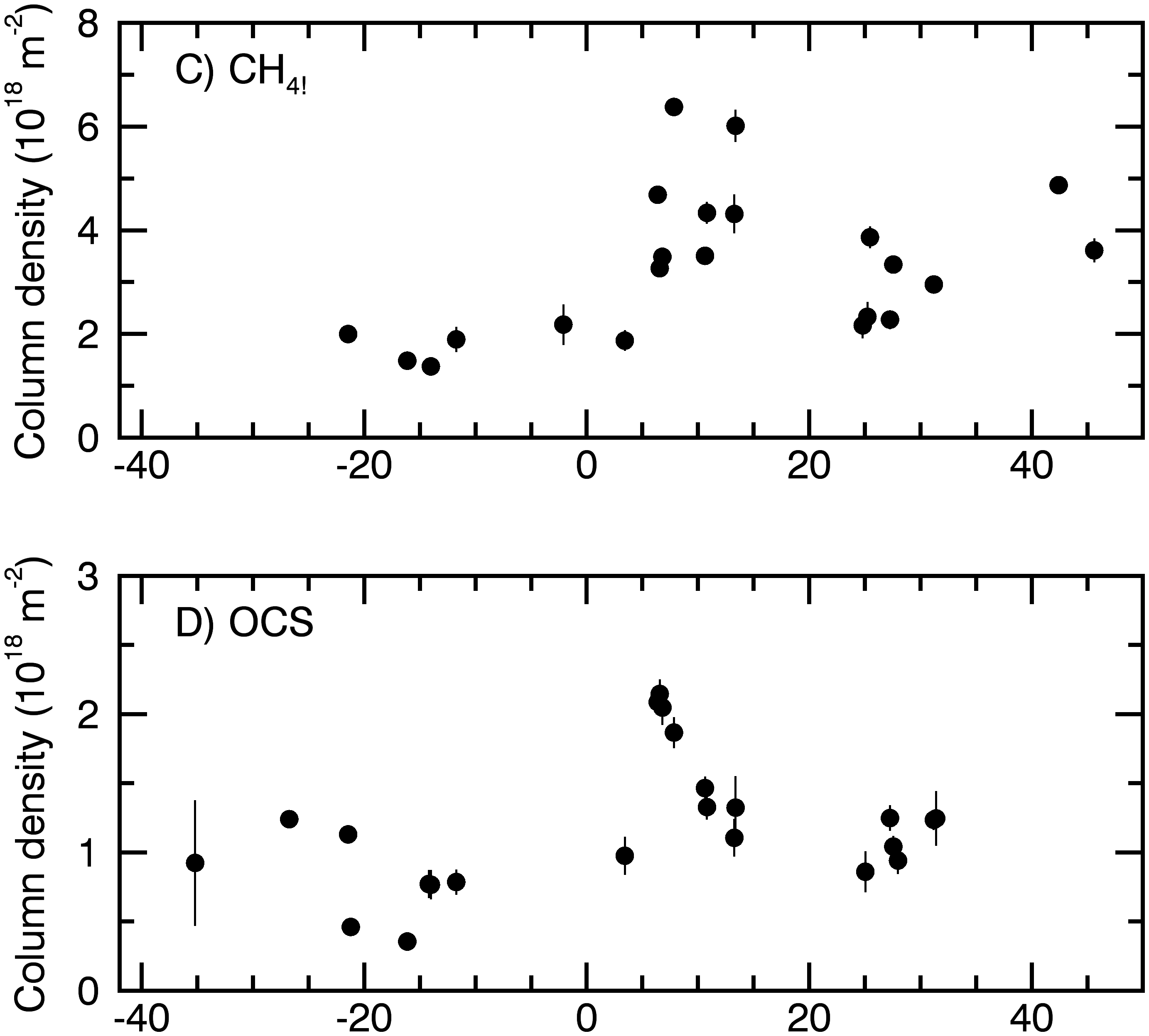}
    \par
    \vspace{0.4cm}
    \includegraphics[width=7.27cm]{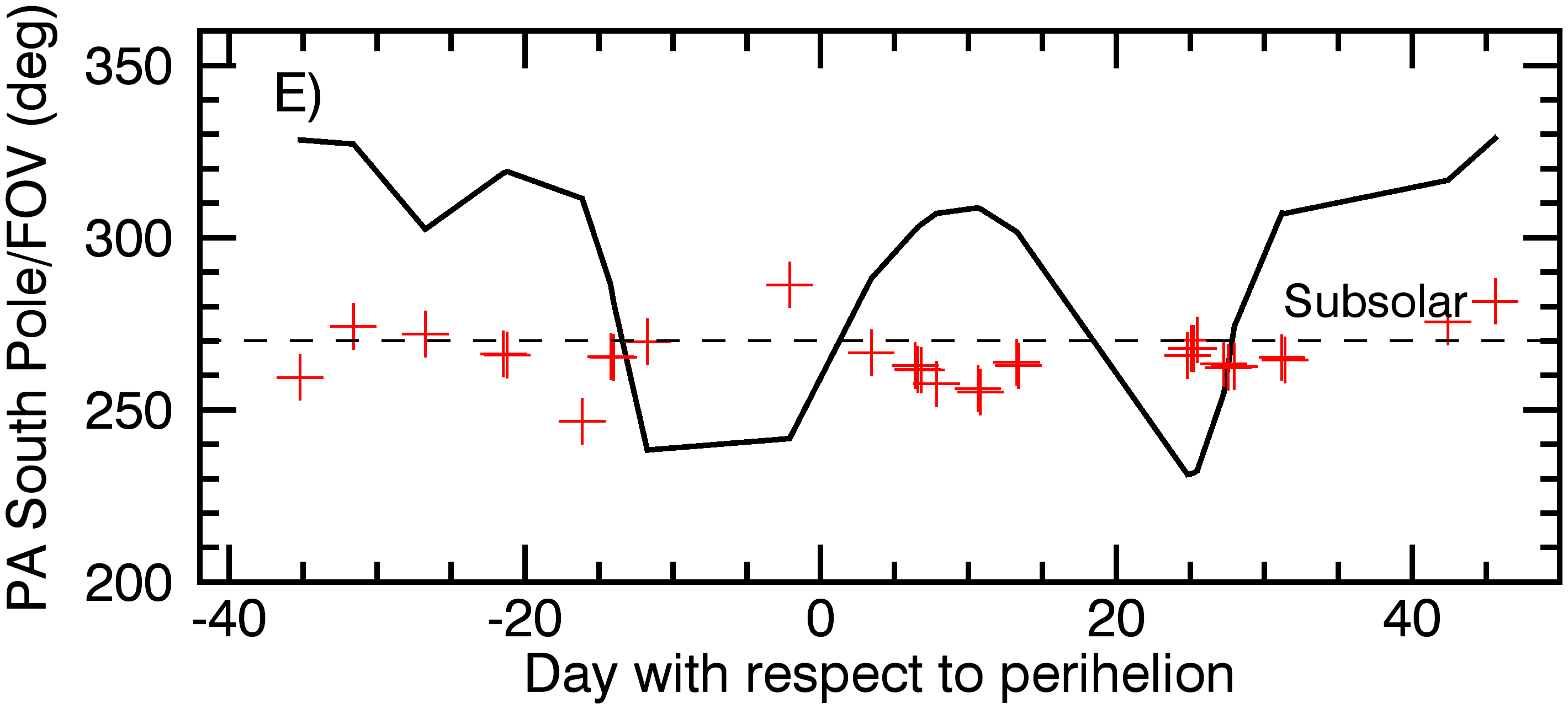}  
   \caption{Column densities of H$_2$O, CO$_2$, CH$_4$ and OCS as a function of date with respect to perihelion (A--D), and position angles of the South pole direction and FOV (E). A--D) Measured column densities (dots with error bar) are rescaled to a distance of $\rho$ = 4 km assuming a 1/$\rho$ variation; the  latitude of the subsolar point is plotted as a dashed line in plot A. E) The PA of the South direction is shown as a black line; the PA of the FOV are plotted with crosses; the projected comet-Sun direction at $PA$ = 270$^{\circ}$ is indicated as a dashed line. }
    \label{fig:H2O-column}
\end{figure}

Band intensities ($I$) and column densities ($N$) obtained from
fitting are listed in Table~\ref{tab:RES-CO2}  for CO$_2$, and in
Table~\ref{tab:RES-CH4} for H$_2$O, OCS, and CH$_4$. Regarding
CO$_2$, we also provide the attenuation factor $f_{atten}$ of the main band which is determined from 
the CO$_2$ and $^{13}$CO$_2$ $\nu_3$ band intensities according to $f_{atten}$ = 89 $\times$ $I$($^{13}$CO$_2$)/$I$(CO$_2$).
The attenuation factor corresponds to the ratio between the $\nu_3$ band intensity expected in optically thin conditions and the observed $\nu_3$ band intensity. This factor reaches values of 2 to 9.
The varying correlation between CO$_2$ band intensity and
attenuation factor is expected from radiative transfer
calculations, because of azimuthal variations of local excitation
related to the uni-directionality of the excitation source and to
the 3D structure of the gas velocity field
\citep{Gersch2014,Debout2016}.

Column densities provide partial information of the overall activity of the comet because: i) the FOV is sampling a small fraction of the coma at a location in the gaseous fan which is each time different (Sect.~\ref{sec:context}); ii) 67P's nucleus shape and outgassing behavior is complex, and we indeed observe rotation-induced variations of column densities (see results for 19 August). The evolution of the water column density in VIRTIS-H field of view is plotted in Fig.~\ref{fig:H2O-column} for $\rho$ = 4 km. We corrected for the different distances from comet centre of the FOV assuming a 1/$\rho$ variation (but we did not correct for possible gas velocity variations). The low H$_2$O column density observed for September 24, 27 (the two last data points) is likely related to the small phase angle, as already discussed in Sect.~\ref{sec:dataset}.  The measured column densities are consistent with those measured with the MIRO instrument (Biver et al. in preparation).

In order to constrain somewhat the CO abundance relative to water from VIRTIS-H spectra, spectral fitting of the 4.45--5.01 $\mu$m region with a model including CO fluorescence emission, in addition to H$_2$O and OCS emissions, has been performed. Retrieved values are between 1 and 2\%, and consistent with Alice and MIRO results (Feldman, personal communication; Biver et al. in preparation). However, we estimate that further analyses on spectra obtained with an improved calibration are required to claim a CO detection in VIRTIS-H spectra.

An increase of water production a few days after perihelion is suggested from the column density, which is not related to the observing geometry (i.e., to a FOV
positionned in a denser part of the gas fan at that time, Fig.~\ref{fig:H2O-column}A). This increase is consistent with the analysis of the VIRTIS-H H$_2$O raster maps obtained regularly from July to November, which show that the water fan had its maximum brightness 6--7 days after perihelion \citep{dbm2016ESLAB}. Therefore, the measured relative abundances should be representative of the composition of the material released from the main outgassing areas in the July--September time frame. Concomitant with the 50\% increase of the water column density, a strong  increase of the column densities (by factors 2--3) of CO$_2$, 
CH$_4$ and OCS is also observed (Fig.~\ref{fig:H2O-column}B--D).

\begin{figure}
    \includegraphics[width=\columnwidth]{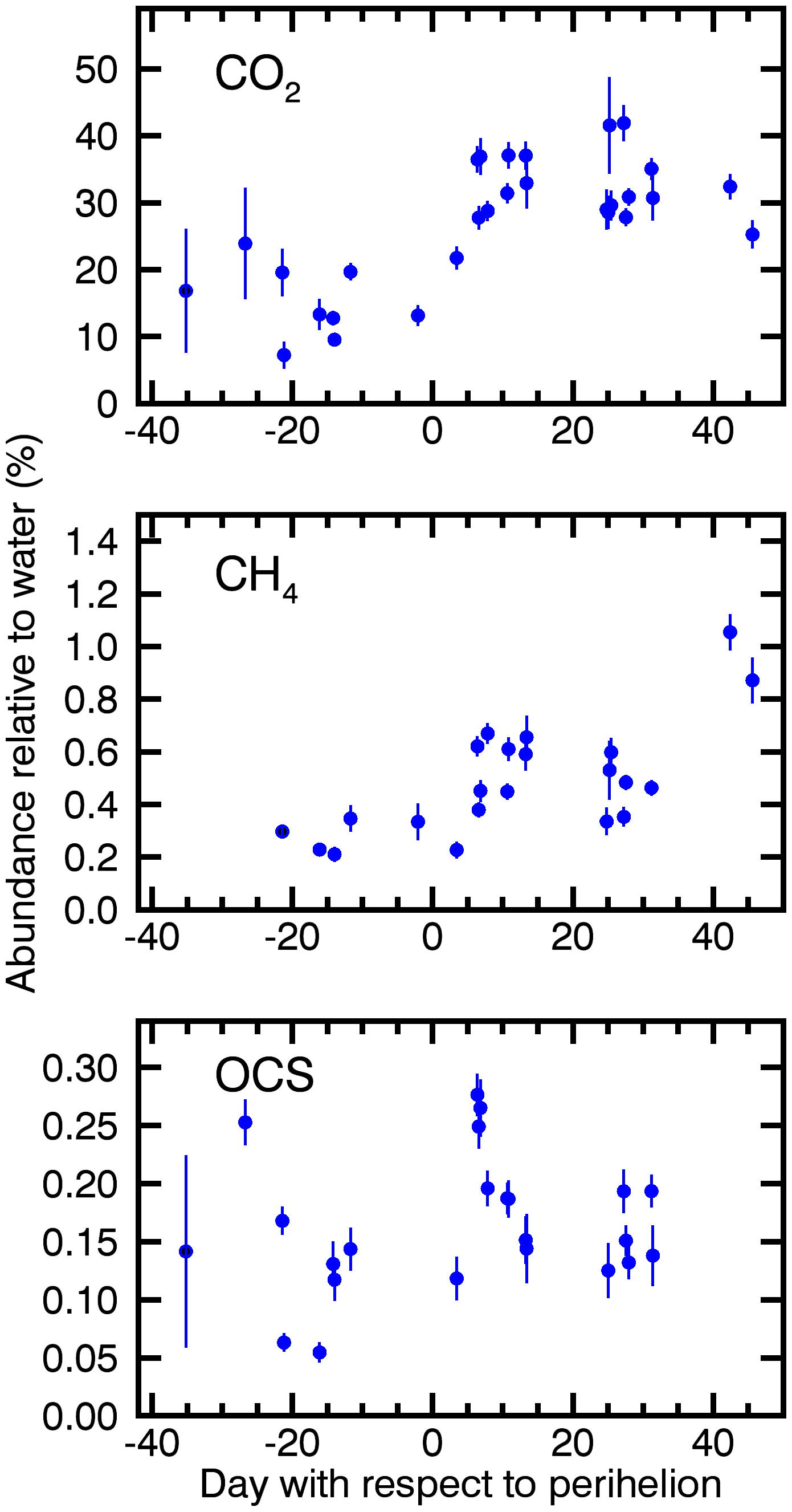}
   \caption{Column density ratios relative to water as a function of day with respect to perihelion.}
    \label{fig:abun}
\end{figure}

Figure \ref{fig:abun} presents column density ratios, with water taken as the reference (hereafter referred to as relative abundances).
A large increase of the CO$_2$ to H$_2$O abundance ratio is measured, starting after perihelion, reaching values in the range 30--40\% from 19 August (i.e., 6 days after perihelion) and thereafter. This increase in the CO$_2$/H$_2$O abundance coincides with the increase in water production. The same trend is observed for CH$_4$, whereas the evolution of the OCS abundance is somewhat chaotic, reaching however its peak value
on 19 August. Averaging data, there is typically a factor of 1.5 increase between OCS abundances measured pre-perihelion and post-perihelion (Table~\ref{tab:abun}). Noteworthy, the investigated post-perihelion period corresponds to the illumination of the most southern latitudes (latitude of the subsolar point $\sim$ --50$^{\circ}$, Fig.~\ref{fig:H2O-column}).

For the discussion in Sect.~\ref{sec:discussion}, we will assume that column density ratios are representative of production rate ratios. This is a good approximation if the spatial distributions of the molecules are similar. However, as judged from the brightness distribution of the CO$_2$ and H$_2$O optically thick bands (Sect.~\ref{sec:context}), the angular width of the CO$_2$ fan is smaller than that of the H$_2$O fan. Therefore, the CO$_2$/H$_2$O production rate ratio from the most active reagions is possibly underestimated.

\begin{table}
\caption{Abundances relative to water.}
\centering
 \label{tab:abun}
    \begin{tabular}{lll}
        \hline
Molecule & Pre-perihelion & Post-perihelion \\
         & 8 Jul.--10 Aug. 2015 &16 Aug.--27 Sep. 2015 \\
         & (\%) & (\%) \\
        \hline
CO$_2$ & 14 & 32 \\
CH$_4$ & 0.23 & 0.47 \\
OCS    & 0.12 & 0.18 \\
    \hline
    \end{tabular}
\end{table}

\section{Discussion}
\label{sec:discussion} The distributions of H$_2$O and CO$_2$ in
the inner coma of 67P have been measured by several Rosetta
instruments since 2014, allowing the investigation of their local,
diurnal and seasonal variations. During pre-equinox, the water
outgassing was found to follow solar illumination, with, however, an
excess of production from the Hapi region in the neck/north pole
area, which photometric properties suggest enrichment in surface
ice \citep{Fornasier2015a,deSanctis2015,Filacchione2016b}. The
CO$_2$/H$_2$O column density ratio above the illuminated Hapi
region was measured to be small. Ratios in the range 1--3\% were
determined from VIRTIS data for the November 2014 to April 2015
period \citep{dbm2015,Migliorini2016}. Similar values were
measured above other regions of the northern hemisphere
\citep{dbm2015,Migliorini2016}. These values are consistent with
measurements at 3 AU from the Sun made by ROSINA 
\citep{Leroy2015}. On the other hand, a large CO$_2$ outgassing
from the non-illuminated southern hemisphere was observed,
indicating a strong dichotomy between the two hemispheres with
subsurface layers much richer in CO$_2$ in the southern regions
\citep{Luspay2015,Migliorini2016,Fink2016}. From annulus measurements surrounding the comet, \citet{Fink2016}
inferred a CO$_2$/H$_2$O total production rate ratio of 3--5 \%
from data obtained in February and April 2015. This ratio is
not diagnostic of the relative amount of gases released from a
specific region, because of the disparate origin of the two molecukes. ROSINA measurements showed that the distribution of polar molecules
is correlated with the H$_2$O distribution (HCN, CH$_3$OH),
whereas that of apolar and more volatile species (CO, C$_2$H$_6$)
is correlated with the CO$_2$ distribution \citep{Luspay2015}.
However, apolar CH$_4$ displayed an intriguing distinct behavior
from both CO$_2$ and water \citep{Luspay2015}. The CH$_4$/CO$_2$
ratio from the northern and southern hemispheres at 3 AU from the
Sun was estimated to be 5 and 0.7\%, respectively from measurements
with ROSINA  \citep{Leroy2015}. As for the CH$_4$/H$_2$O ratio it
was estimated to 0.1\% above the northern summer hemisphere
\citep{Leroy2015}. At 3 AU from the Sun, 67P was releasing small
amounts of OCS: OCS/H$_2$O of 0.02\% above the northern hemisphere
and OCS/CO$_2$ of 0.1 \% above the southern hemisphere.

In this paper, we investigated outgassing from the southern
regions when illuminated by the Sun. These regions were clearly
much more productive than during the southern winter time. Using column
densities at 4 km from the surface as a proxy of activity,
$N$(CO$_2$) increased typically from $\sim$ 10$^{19}$ m$^{-2}$  at
$r_h$ = 1.9 AU \citep{Migliorini2016} to a few 10$^{20}$ m$^{-2}$
at perihelion time (Table~\ref{tab:RES-CO2}). The CH$_4$/CO$_2$
ratio remained within a factor of 2 the value measured at 3 AU,
whereas the OCS/CO$_2$ ratio increased by a factor of 6--9.
Comparing now the relative  amounts (with respect to water) of
gases released from the two hemispheres when illuminated (keeping
in mind the different heliocentric distances), the southern
hemisphere is up to a factor of 10 more productive in CO$_2$
and OCS relative to water, and by a factor of 2--5 more productive
in CH$_4$. Table ~\ref{tab:abun-seasons} summarizes relative
abundances measured above the two hemispheres at different
seasons.

This seasonal variation in relative molecular abundances is
related to the strong differences in illumination conditions
experienced by the two hemispheres. The northern hemisphere
experiences a long summer but at large heliocentric distances,
implying low sublimation rates and very low surface ablation.
Though 67P surface and subsurface layers present a low thermal
inertia, a stratification in the ice composition is expected, with
the more volatile species residing in deeper layers. The residence
level of the volatile ices may be at a few tens centimeters from
the surface, down to several meters, depending on the volatility of
the ice (CO$_2$, OCS, CH$_4$ in increasing order of volatility),
the dust mantle thickness, thermal inertia, porosity, and
illumination history \citep{Marboeuf2014}. The detection of
low levels of CO$_2$ production from the northern hemisphere
indicates that probably most of the CO$_2$ ice was below the level
of the thermal wave, with only a small portion of CO$_2$ ice
reached by the heat. Based on modelling, the CO$_2$/H$_2$O ratio
measured in the northern hemisphere during summer (1--3 \%) should
be lower than the nucleus undifferentiated  value
\citep{Marboeuf2014}. In other words, values measured in the
northern hemisphere are not original, but the result of the
devolatilization of the uppermost layers.

The increased gaseous activity of comet 67P as it approached the
Sun led to the removal of devolatilized dust, thereby exposing
less altered material to the Sun. Evidence for rejuvenation of the
surface of 67P is given by a significant increase of the single
scattering albedo and more bluish colors as 67P  approached the
Sun \citep{Filacchione2016b,Ciarnello2016}. The absence of wide
spread smooth or dust covered terrains in the southern hemisphere,
together with the presence of large bright ice-rich areas
\citep{Elmaarry2016,Fornasier2015b} are lines of evidence that this hemisphere
was subjected to strong ablation during the perihelion period.
Models of the thermal evolution of 67P predict that the depth of
ablation should have reached the underlayers where volatile ices
are present \citep{deSanctis2010b,Marboeuf2014}. The observed
enhancement of the CO$_2$/H$_2$O, CH$_4$/H$_2$O, and OCS/H$_2$O
abundance ratios could be the result of this ablation. Once the removal
of devolatilized layers has proceeded, sustained dust removal due
to the sublimation of water ice maintained volatile-rich layers
near the surface, as suggested by the abundance ratios which remained high long after
perihelion (Fig.\ref{fig:abun}). This requires the erosion velocity to be 
comparable or larger than the propagation velocity of the diurnal heat, a property expected near perihelion for low thermal inertia material \citep{Gortsas2011}. Should ablation have reached
non-differentiated layers, abundances relative to water
measured in the coma would then be representative of the nucleus
initial ice composition. Simulations performed for comet 67P by \citet{Marboeuf2014} (though not considering the large obliquity of the spin axis) indeed predict a CO$_2$/H$_2$O production rate ratio near perihelion similar to the primitive composition of the nucleus for active regions not covered by a thick dust mantle. 
This model includes diurnal and seasonal recondensation of gases in the colder layers. A result of  these simulations is that this process is ineffective in altering the CO$_2$/H$_2$O production rate ratio from the southern hemisphere due to the strong erosion. Further modelling of 67P activity may provide alternative interpretations.

OCS is only slightly more volatile than CO$_2$ with a sublimation temperature
of 57 K to be compared to 72 K for the latter species \citep{Yamamoto1985}. Therefore we should expect similar production curves, if these species are present in icy form. The mean OCS/CO$_2$ ratios measured pre and post-perihelion in the 7 July to 27 September 2015 period are effectively similar (0.6--0.9 \%, Table~\ref{tab:abun-seasons}) and comparable to the value measured above the northern hemisphere at 3 AU from the Sun by ROSINA \citep{Leroy2015}.  However, the OCS/CO$_2$ ratio measured above the southern hemisphere at 3 AU from the Sun is 7 times lower \citep{Leroy2015}. Possibly, this means that the thermal wave did not reach OCS-rich layers. Also, looking to the individual VIRTIS data obtained in the perihelion period, fluctuations in the OCS/CO$_2$ ratio are observed, possibly related to compositional or structural heterogeneities.  A better understanding of the differences in OCS and CO$_2$ outgassing would need a larger data set (as that obtained by ROSINA).

As in the case of CO$_2$, the observed large increase in CH$_4$ column density and CH$_4$/H$_2$O ratio at perihelion is consistent with a CH$_4$ outgassing front near the surface. The evolution of the sublimation interface of CH$_4$ pure ice, which is more volatile than CO$_2$ ice, has been examined by \citet{Marboeuf2014}. In this model, the ablation does not reach the CH$_4$ ice front, and no significant enhancement in CH$_4$ production is expected at perihelion. However, a large increase in CH$_4$ production would occur for CH$_4$ trapped in amorphous ice as the ablation reaches the crystallisation front, or for CH$_4$ encaged in a clathrate hydrate structure. In the model of \citet{Marboeuf2014}, an obliquity of 0$^{\circ}$ is assumed, and results are provided for equatorial latitudes. Results obtained considering the 52$^{\circ}$ obliquity of 67P spin axis are given by \citet{deSanctis2010b} for CO and CO$_2$ present in icy form. Contrary to \citet{Marboeuf2014}, this model predicts a strong outgassing of both CO$_2$ and CO in the south hemisphere at perihelion. This is because the important erosion that takes place in these regions during their short summer maintains the sublimation interface of volatile species as CO and CH$_4$ close to the surface.

The CH$_4$/H$_2$O ratio exhibits much less contrasting regional and seasonal variations than observed for CO$_2$ (Table~\ref{tab:abun-seasons}). Howewer, the same trend is observed. The maximum value is for South/winter, the minimum value is for North/summer, whereas the intermediate value is for South/summer. This might suggest that they are in a similar form and located in layers close to each other. Should CH$_4$ and CO$_2$ be both present in icy form, then less contrasted variations are expected for CH$_4$ because of its smaller sublimation temperature. Alternatively, CH$_4$ could be trapped in amorphous water ice inside 67P nucleus and released during ice crystallization, which occurs at a temperature a little below that of water ice sublimation. This could explain the small difference in CH$_4$/H$_2$O ratios measured above the two
hemispheres during summer time. However, this scenario cannot explain the significant CH$_4$ outgassing from the non-illuminated hemisphere at 3 AU from the Sun \citep{Luspay2015}. \citet{Luspay2016} propose the presence of CH$_4$ clathrate hydrates to explain the behavior observed by ROSINA.     

The column densities of H$_2$O, CO$_2$, CH$_4$ and OCS show an abrupt increase at about 6 days after perihelion (Fig.~\ref{fig:H2O-column}). This is an interesting coincidence with the results of the thermal model of \citet{Keller2015} applied to the real shape of 67P, which lead to a time lag of 6 days caused by the asymmetry of the subsolar latitude variation with respect to perihelion. However, this predicted maximum is quite flat and does not fit to the sudden increase we measured. An alternative explanation could be that the sublimation front is at some depth $x$ below the surface and is reached by the maximum insolation energy of perihelion with some delay. The timescale $\tau$ of thermal conduction is defined by \citep{Gortsas2011}:

\begin{equation}
\tau  = \rho  c x^2 / k, 
\end{equation}

\noindent
where $\rho$ and c are the density and specific heat of the surface material, respectively, and $k$ is its thermal conductivity. Using $\tau$ = 6 days ($\sim$ 500000 s) and standard thermo-physical parameters for 67P \citep{Shi2016} of $\rho$ = 500 kg/m3 , $c$ = 600 J/kg K, and $k$ = 0.005 W/mK one gets a depth of about 7 cm. But also in this model one would expect a gradual increase of the production rates after perihelion. It is interesting to note that all measured species show nearly the same time lag although having different volatilities. According to \citet{Marboeuf2014}, Fig. 4, none of the studied outgassing profiles (amorphous, clathrate, crystalline, mixed) shows a similar behavior as measured here. A plausible explanation could be that a volatile poor insulating surface layer is quickly eroded near perihelion leading to a strong increase of activity of several volatiles in short time.

Production rate ratios relative to water of CO$_2$, OCS and CH$_4$
have been measured in numerous comets, and display variations by a
factor of up to $\sim$ 10 (Table~\ref{tab:abun-seasons}). The
CO$_2$/H$_2$O ratio measured post-perihelion in 67P indicates that
this comet is CO$_2$--rich. The OCS/H$_2$O ratio is in the range
of values measured in other comets, though the post-perihelion
value is slightly above the mean value of $\sim$ 0.1\%
\citep{Biver2016}. The CH$_4$/H$_2$O ratio in 67P is also within
the range the values measured in other comets, but is by a factor
2 lower than the 'typical' value of $\sim$ 1\% \citep{Mumma2011}.

The VIRTIS-H observations are consistent with a small CO/H$_2$O production rate ratio (a few percents at most) from the southern hemisphere. This is in the range of values measured for Jupiter-family comets (JFC) \citep[see, e.g.,][]{Ahearn2012}. The number of comets for which both CO and CO$_2$ abundances relative to water have been measured is very limited. Among the few JFC comets of the sample, only the small lobe of 103P/Hartley 2 investigated by the EPOXI mission have similarities with 67P southern hemisphere, that is a high CO$_2$/H$_2$O ratio and a low CO/H$_2$O ratio. The CO/CO$_2$ ratio is 0.01 \citep[][and references therein]{Ahearn2012}, to be compared to the value of $\sim$ 0.03 measured above 67P southern hemisphere, assuming CO/H$_2$O = 1 \% (Biver et al., personal communication).

\section{Conclusion}

This paper focussed on the perihelion period during
which the southern regions of comet 67P were heavily illuminated.
The measured abundance ratios of CO$_2$, CH$_4$ and OCS, and reported trends show that
seasons play an important role in comet outgassing. The  
high CO$_2$/H$_2$O ratio measured in the coma is plausibly reflecting the pristine
ratio of 67P nucleus, indicating a CO$_2$ rich comet. On the other hand, the low CO$_2$/H$_2$O values 
measured above 
the illuminated northern hemisphere are most likely not original, but the
result of the devolatilization of the uppermost surface layers. The VIRTIS-H
instrument acquired infrared spectra of H$_2$O and CO$_2$ during
the whole Rosetta mission. Analysis of the whole data set will
certainly provide further information on outgassing processes
taking place on 67P's nucleus.

\begin{table}
\caption{Seasonal variations of abundance ratios (in \%) measured in 67P.}
\centering
 \label{tab:abun-seasons}
    \begin{tabular}{lcccc}
        \hline
Ratio & North & \multicolumn{2}{c}{South} & Other comets$^d$\\
(\%)  & summer$^a$ & winter$^b$ & summer$^c$ & \\
        \hline
CO$_2$/H$_2$O & 1--3 & 80 &14--32  & 2.5--30 \\
CH$_4$/H$_2$O & 0.13 & 0.56 & 0.23-0.47 &  0.12--1.5 \\
OCS/H$_2$O    & 0.017 & 0.098 &  0.12-0.18  & 0.03--0.4\\
\hline
CH$_4$/CO$_2$ & 5 & 0.7 & 1.5--1.6 \\
OCS/CO$_2$  & 0.7 & 0.1 &  0.6--0.9 \\
    \hline
    \end{tabular}

            {\raggedright
    $^a$ Values at $r_h$ = 3 AU from \citet{Leroy2015}, except for CO$_2$ measured at 1.8--2.9 AU \citep{dbm2015,Migliorini2016}.

    $^b$ Values at $r_h$ = 3 AU from \citet{Leroy2015}.

    $^c$ Values from this work at $r_h$ = 1.25--1.36 AU.

    $^d$ \citet{Biver2016}, \citet{Ootsubo2012}, \citet{McKay2016}, \citet{Mumma2011}.

    }
\end{table}

\section*{Acknowledgements}
The authors would like to thank the following institutions and
agencies,  which supported this work: Italian Space Agency (ASI -
Italy), Centre National d'Etudes Spatiales (CNES- France),
Deutsches Zentrum f\"{u}r Luft- und Raumfahrt (DLR-Germany),
National Aeronautic and Space Administration (NASA-USA). VIRTIS
was built by a consortium from Italy, France and Germany, under
the scientific responsibility of the Istituto di Astrofisica e
Planetologia Spaziali of INAF, Rome (IT), which lead also the
scientific operations.  The VIRTIS instrument development for ESA
has been funded and managed by ASI, with contributions from
Observatoire de Meudon financed by CNES and from DLR. The
instrument industrial prime contractor was former Officine
Galileo, now Selex ES (Finmeccanica Group) in Campi Bisenzio,
Florence, IT. The authors wish to thank the Rosetta Science Ground
Segment and the Rosetta Mission Operations Centre for their
fantastic support throughout the early phases of the mission. The
VIRTIS calibrated data shall be available through the ESA's
Planetary Science Archive (PSA) Web site. With fond memories of
Angioletta Coradini, conceiver of the VIRTIS instrument, our
leader and friend. D.B.M. thanks G. Villanueva for providing H$_2$O
fluorescence calculations.

\bsp	
\label{lastpage}
\end{document}